\documentclass[reprint,superscriptaddress,english,preprintnumbers,amsmath,amssymb,aps,prx]{revtex4-2}
\usepackage{graphicx}
\usepackage{dsfont}
\usepackage{xcolor}
\usepackage{multirow}
\usepackage{breakurl}
\usepackage[normalem]{ulem}
\usepackage{enumitem} 
\usepackage{charter} 
\usepackage[charter]{mathdesign}

\definecolor{refcol}{RGB}{178,34,34}
\usepackage{microtype}
\usepackage[colorlinks,linkcolor=refcol,citecolor=refcol,urlcolor=refcol]{hyperref}

\graphicspath{{./figures/}}

\def\det{{\mathrm{det}}}

\def\eq#1{(\ref{#1})}

\def\Eq#1{Eq.~(\ref{#1})}

\def\Fig#1{Fig.~\ref{#1}}

\def\Sec#1{Sec.~\ref{#1}}
\def\App#1{App.~\ref{#1}}

\definecolor{blue}{rgb}{0,0,1}

\definecolor{green}{rgb}{0,1,0}

\definecolor{red}{rgb}{1,0,0}

\begin{document}

\title{Particle Interferometry in a Moat Regime}

\author{Fabian Rennecke}
\email{fabian.rennecke@theo.physik.uni-giessen.de} 
\affiliation{Institute for Theoretical Physics, Justus Liebig University Giessen, Heinrich-Buff-Ring 16, 35392 Giessen, Germany}
\affiliation{Helmholtz Research Academy Hesse for FAIR (HFHF), Campus Giessen, Giessen, Germany}

\author{Robert D.\ Pisarski}
\email{pisarski@bnl.gov}
\affiliation{Department of Physics, Brookhaven National Laboratory, Upton, New York 11973, USA}

\author{Dirk H.\ Rischke}
\email{drischke@itp.uni-frankfurt.de}
\affiliation{Institute for Theoretical Physics, Goethe University, Max-von-Laue-Str.\ 1, 60438 Frankfurt am Main, Germany}
\affiliation{Helmholtz Research Academy Hesse for FAIR (HFHF), Campus Riedberg, Frankfurt am Main, Germany}

\begin{abstract}
Dense strongly interacting matter can exhibit regimes with spatial modulations, akin to crystalline phases. In this case particles can have a moat spectrum with minimal energy at nonzero momentum. We show that particle interferometry is a sensitive probe of such a regime in heavy-ion collisions. To this end,   
we develop a field-theoretical formalism 
that relates particle spectra to in-medium real-time correlation functions of quantum fields on curved hypersurfaces of spacetime.
This is then applied to the study of Bose-Einstein correlations in a moat regime in heavy-ion collisions. The resulting two-particle spectra exhibit peaks at nonzero average pair momentum, in contrast to the
two-particle spectra in a normal phase, which peak at zero momentum. These peaks lead to non-trivial
structures in the ratio
of two-particle correlation functions, which should be experimentally measurable if the resolution in the direction of average pair momentum is sufficiently large. We propose these structures in the correlation-function ratios as clear signature of a moat regime and spatially modulated phases in quantum chromodynamics (QCD). 
\end{abstract}

\maketitle


\section{Introduction}\label{sec:intro}

 Understanding the phase structure of QCD is a long-standing open problem of fundamental physics. Lattice QCD \cite{Aoki:2006we, Borsanyi:2010bp,HotQCD:2018pds} and functional continuum methods \cite{Fischer:2018sdj,Fu:2019hdw,Gao:2020fbl} have by now established that QCD at vanishing and small densities has a crossover transition from hadrons at low temperature to deconfined quarks and gluons at very high temperature. At intermediate temperatures, matter could be in a `stringy fluid' state \cite{Glozman:2022lda}. For extremely high densities, well above any density found in Nature, it is known that at low temperatures QCD is in a color-superconducting state \cite{Son:1998uk, Pisarski:1999tv, Alford:2007xm}. At intermediate densities there are presently no first-principles results available, but a wealth of different phases have been predicted by model studies \cite{Fukushima:2010bq}.

Strongly interacting matter in this intermediate density region is of particular interest, as it is produced in heavy-ion collisions and is present in neutron stars \cite{Lovato:2022vgq}. Among the phases suggested by model studies are such with a spatial modulation of a certain quantity, like the density or the order parameter for chiral symmetry breaking or color superconductivity. These could be crystalline or inhomogeneous phases \cite{Fukushima:2010bq, Buballa:2014tba},  liquid-crystal like phases \cite{Lee:2015bva, Hidaka:2015xza, Pisarski:2018bct}, or a quantum pion liquid \cite{Pisarski:2020dnx,Pisarski:2021aoz}. Where and if at all these phases are realized in the phase diagram is ultimately decided by long-range quantum fluctuations, and understanding them therefore poses a great challenge. However, a characteristic, or at least precursory, feature of all these phases is the occurrence of a \emph{moat} energy spectrum \cite{Pisarski:2020dnx,Pisarski:2021qof,Rennecke:2021ovl}. In a low-momentum expansion, the energy of a relativistic particle (assuming rotational invariance) is
\begin{align}\label{eq:Eq}
E_{\bold{p}} = \sqrt{z\,\bold{p}^2 + w\,\bold{p}^4 + m^2}\,.
\end{align}
A free relativistic particle has $z=1$ and $w=0$ and thus a minimal energy at rest, $\bold{p}_{\rm min} = 0$. In contrast, in the presence of spatial modulations with a wavenumber $\bold{k}_0 \neq 0$, the energy of particles is minimized at nonzero momentum, $\bold{p}_{\rm min} = \bold{k}_0$. At low momenta, their energy is described by \Eq{eq:Eq} with $z<1$ and, for stability, $w>0$. The energy then looks like a moat. We refer to the region of the phase diagram where particles have such an energy as a \emph{moat regime}.

Since a moat regime occurs whenever spatial modulations are present, but is to some extent agnostic to which type of phase is actually realized, it is both theoretically and, as we will show here, also experimentally more easily accessible. The moat regime can be viewed as an indicator for the presence of spatially modulated phases in QCD. First theoretical evidence that this is indeed the case has been found in Ref.\ \cite{Fu:2019hdw}. Stimulated by the first suggestions in Refs.\ \cite{Pisarski:2020gkx, Pisarski:2021qof}, we will show how a moat regime can be detected in heavy-ion collisions.

\begin{figure}[t]
\centering
\includegraphics[width=.9\columnwidth]{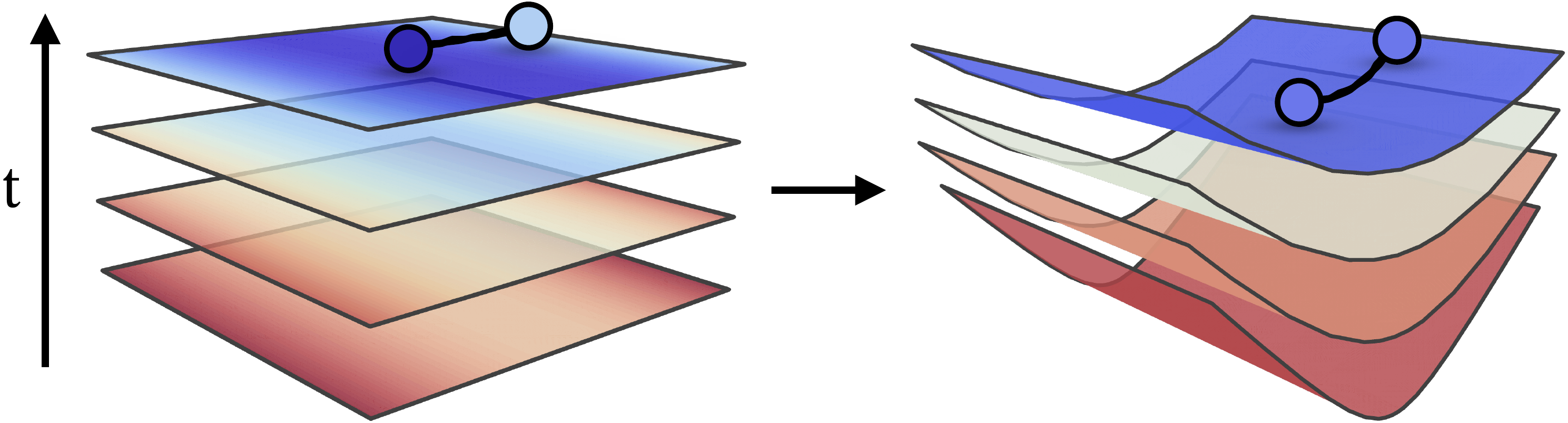}
\caption{Instead of describing correlations at equal time on a flat hypersurface of spacetime, where the involved particles can be subject to different thermodynamic conditions (left figure), we describe them on a curved hypersurface defined by fixed thermodynamic conditions (right figure).}
\label{fig:foliation}
\end{figure}

\begin{figure*}[t]
\centering
\includegraphics[width=.32\textwidth]{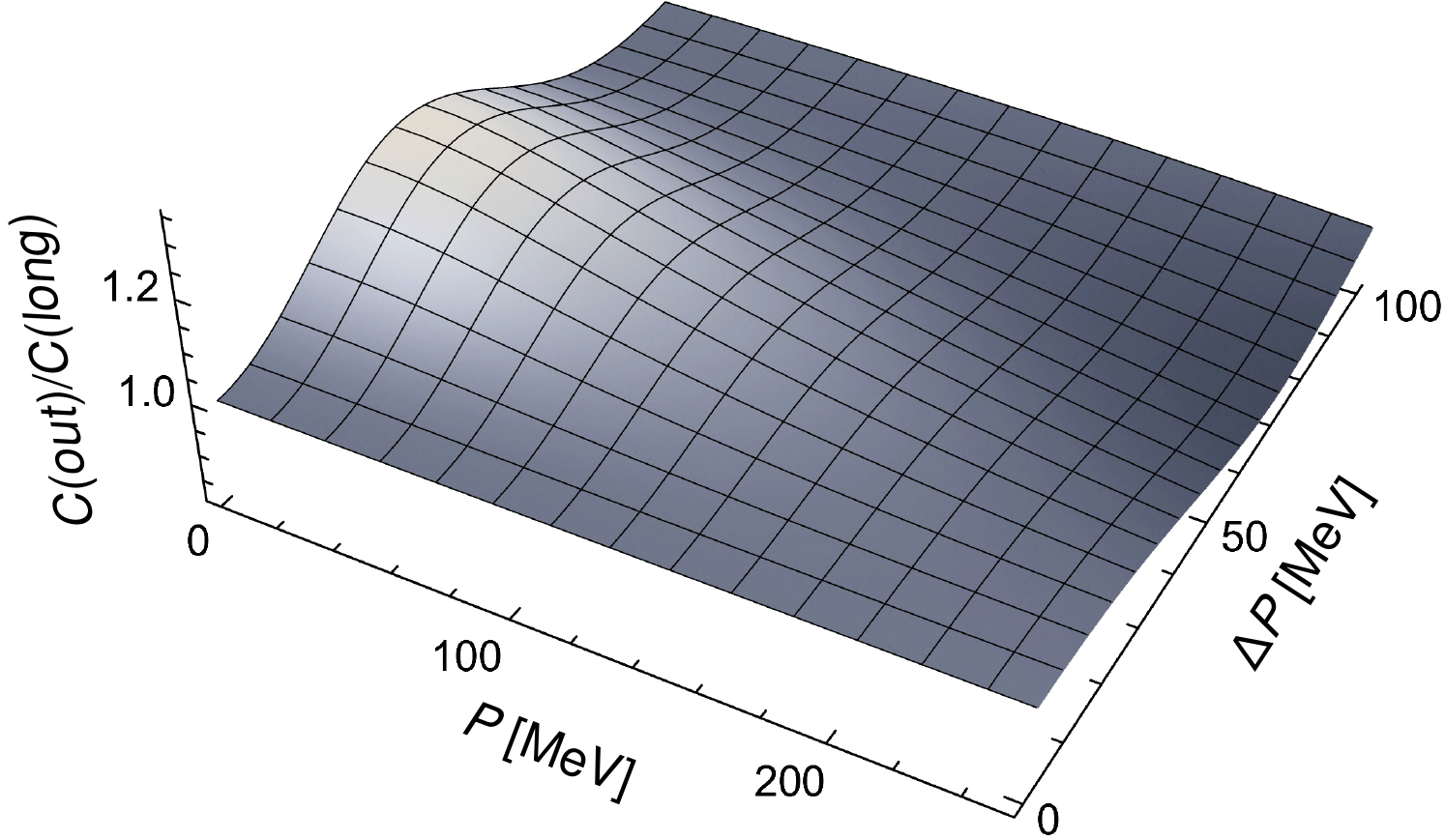}
\hfill
\includegraphics[width=.32\textwidth]{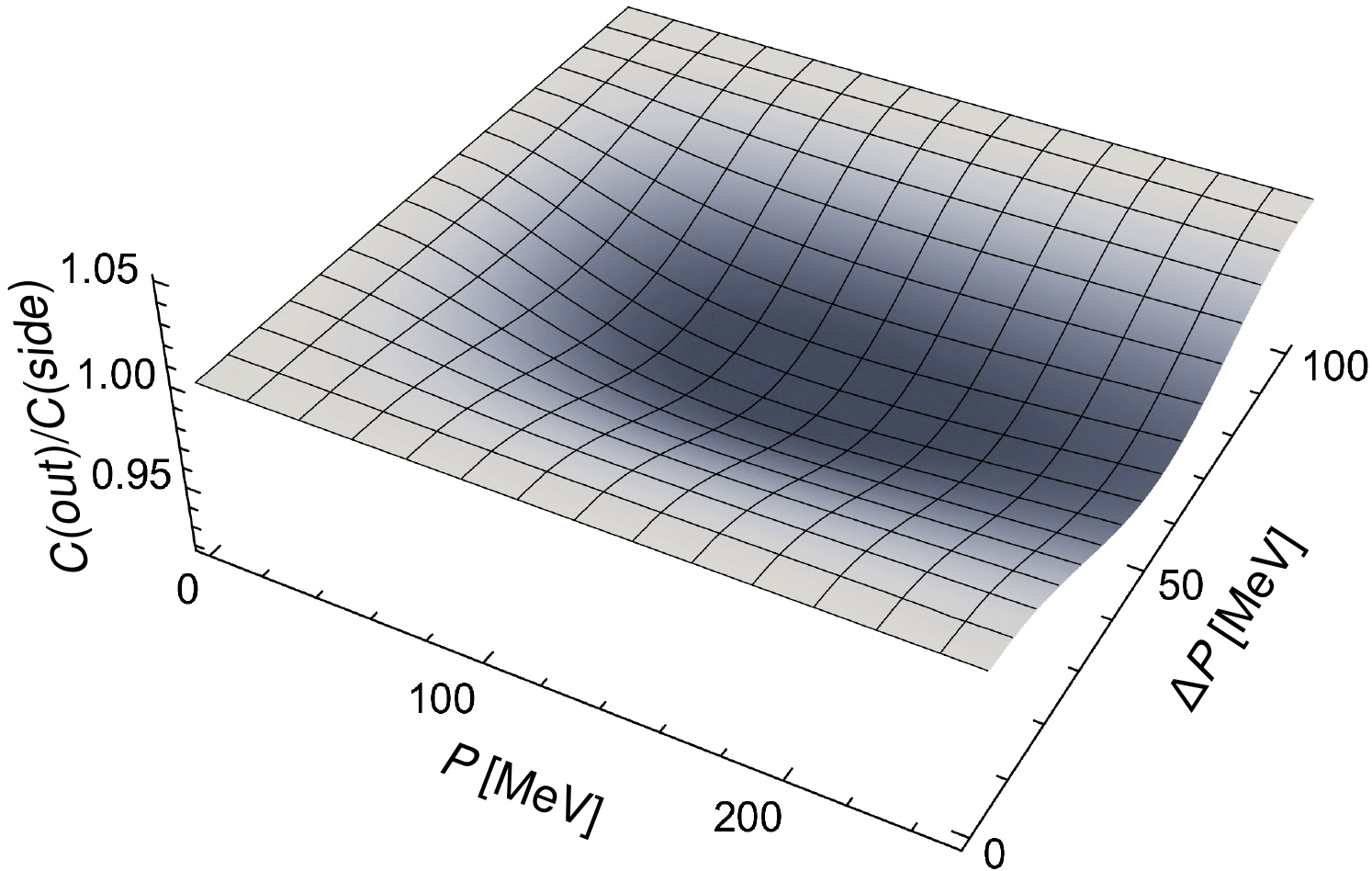}
\hfill
\includegraphics[width=.32\textwidth]{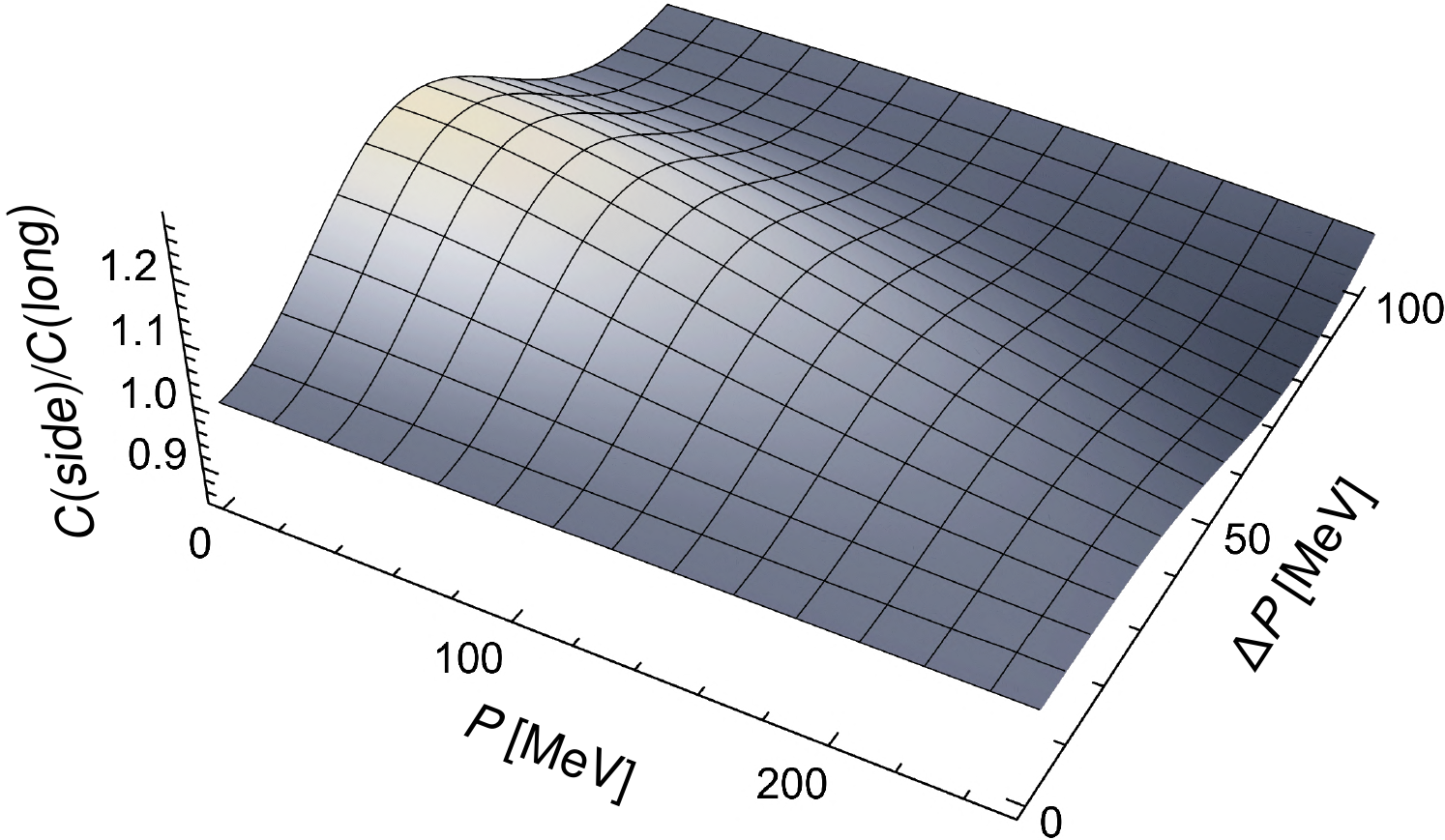}
\caption{Ratios of normalized two-particle correlations, $\frac{C(P,\Delta P_{\rm out})}{C(P,\Delta P_{\rm long})}$, $\frac{C(P,\Delta P_{\rm out})}{C(P,\Delta P_{\rm side})}$ and $\frac{C(P,\Delta P_{\rm side})}{C(P,\Delta P_{\rm long})}$ in the \emph{normal phase}. \textit{Out}, \textit{side}, and \textit{long} denote the direction of the relative momentum of the particle pair, $\Delta P = |\bold{\Delta P}|$, relative to the average pair momentum $P = |\bold{P}|$. 
In the standard notation, the average pair momentum $P \to K$, and
the relative momentum $\Delta P \to q$.}
\label{fig:CratConv}
\end{figure*}

\begin{figure*}[t]
\centering
\includegraphics[width=.32\textwidth]{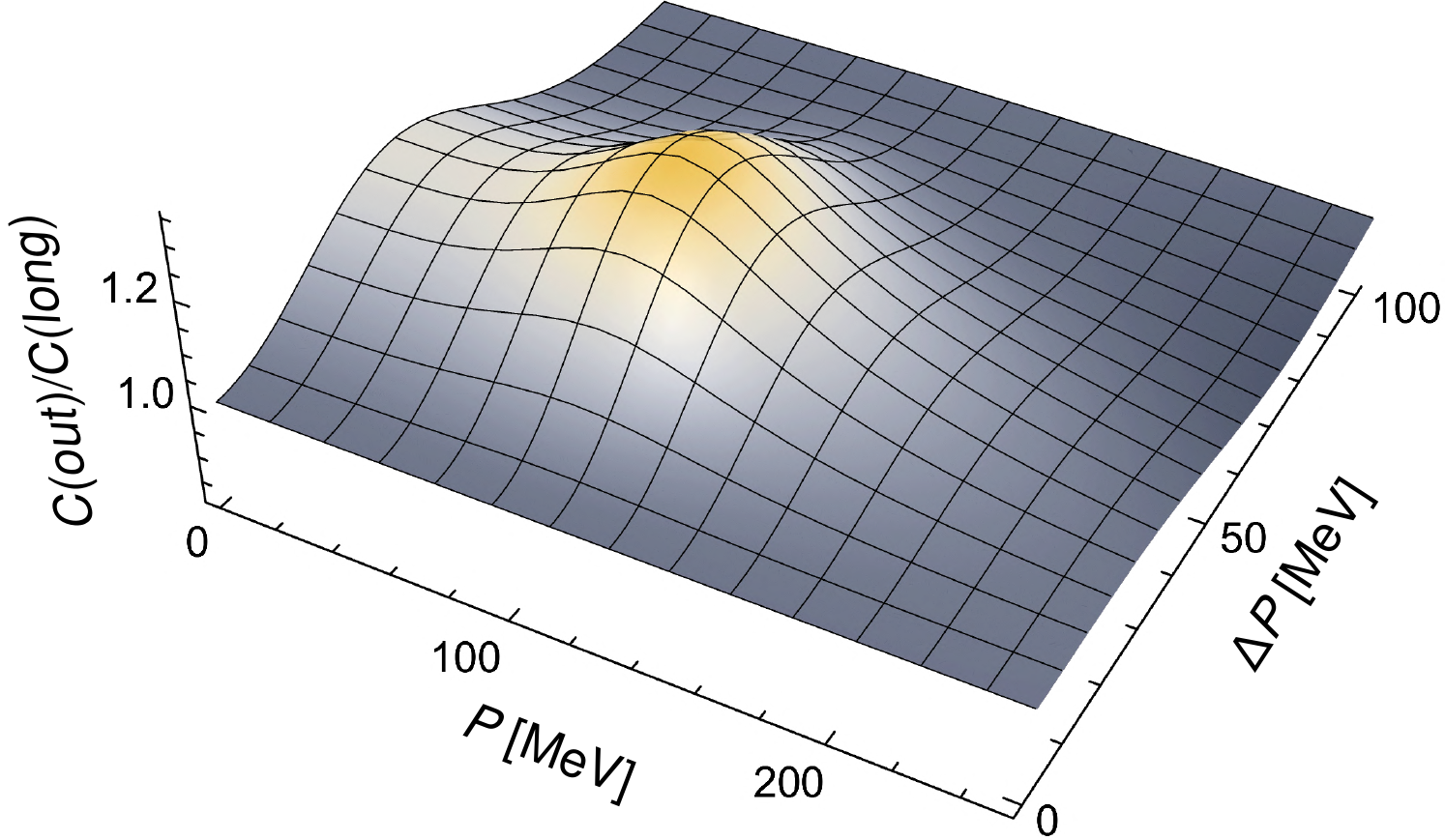}
\hfill
\includegraphics[width=.32\textwidth]{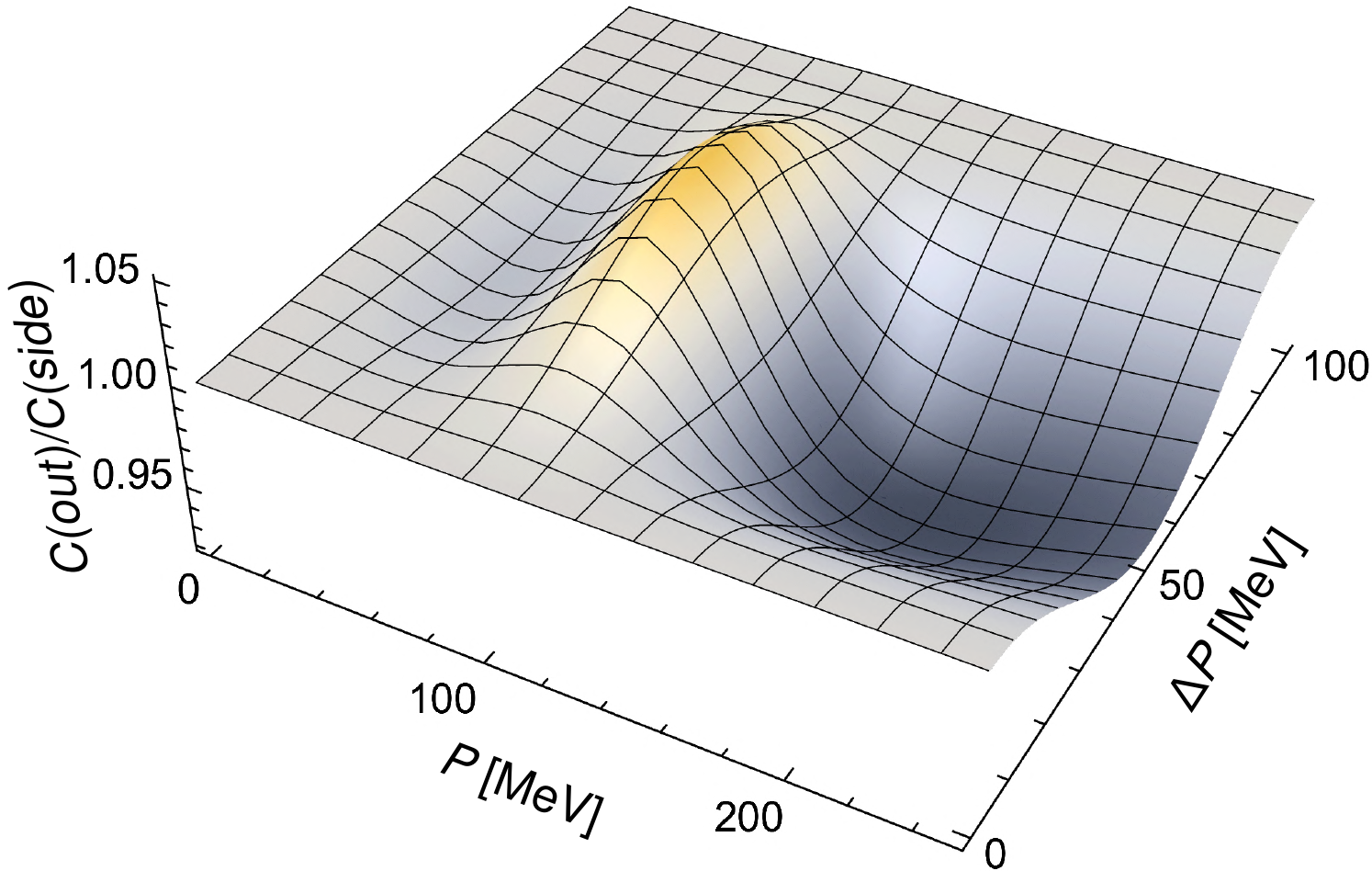}
\hfill
\includegraphics[width=.32\textwidth]{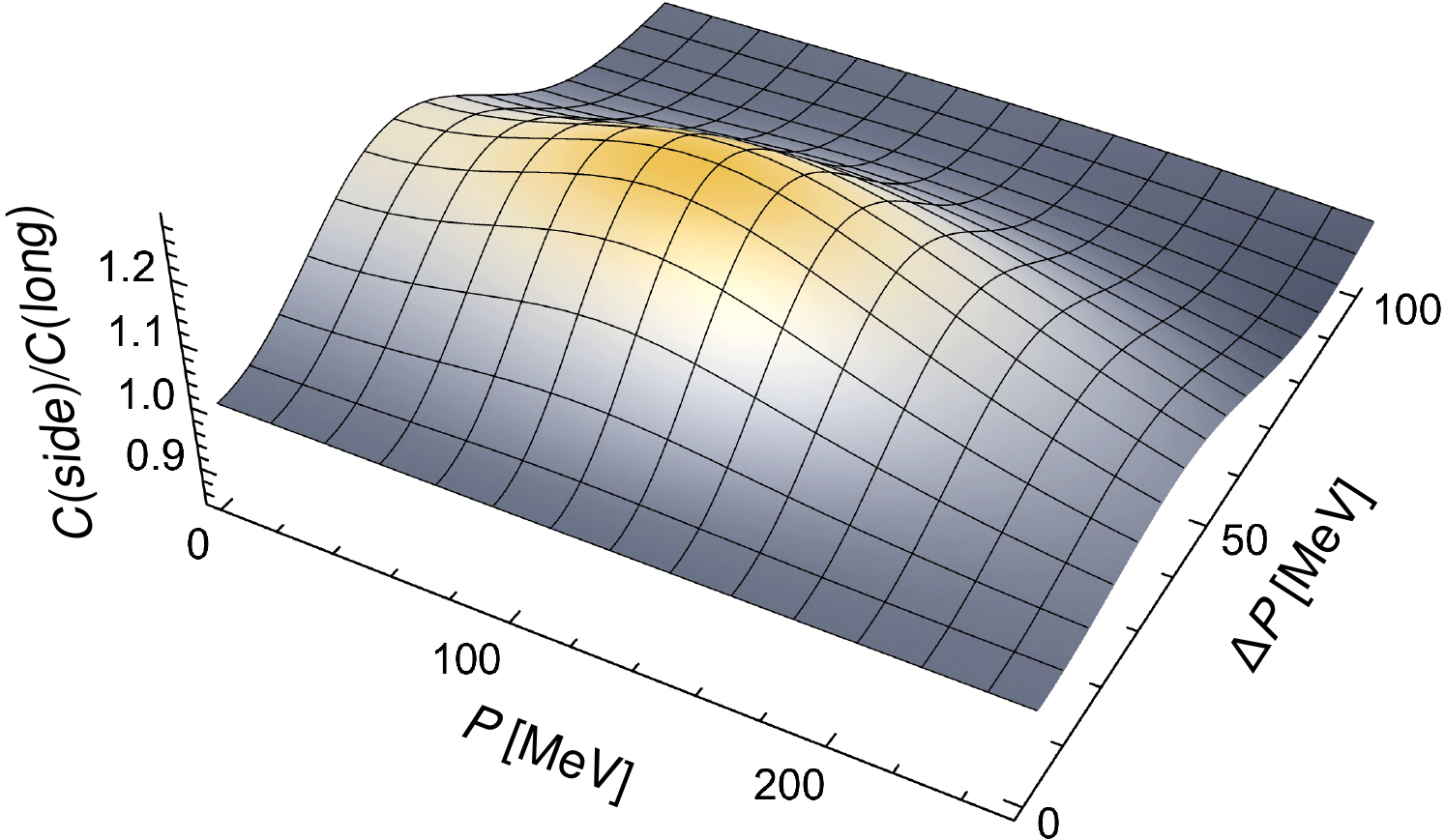}
\caption{The same ratios as in \Fig{fig:CratConv}, but in a \emph{moat regime}. The additional peaks around $P \approx 100$\,MeV reflect the wave number of the underlying spatial modulation and are therefore clear signatures of a moat regime.}
\label{fig:CratMoat}
\end{figure*}

The basic idea is that since particles in a moat regime have minimal energy at nonzero momentum $\bold{p}_{\rm min}$, particle production is enhanced at this momentum. It has been shown in Ref.\ \cite{Pisarski:2021qof} that this leads to a peak in the single-particle spectrum which is directly related to $\bold{p}_{\rm min}$, and that such a peak persists in correlations that are generated by thermodynamic fluctuations. Another prominent source of correlations, which so far has not been studied in this context, is due to the quantum statistics of identical particles. The study of these correlations is known as Hanbury-Brown--Twiss (HBT) interferometry, and was originally used to infer the angular diameter of astronomical objects \cite{HanburyBrown:1956bqd}. In particle physics, HBT correlations through interference measurements of pions were first studied in Ref.\ \cite{Goldhaber:1960sf}. Since then, identical-particle interferometry has been developed into the standard tool for the extraction of information on the spacetime structure of the medium created in heavy-ion collisions \cite{Heinz:1996bs, Heinz:1999rw, Wiedemann:1999qn, Weiner:1999th, Lisa:2005dd}.

Since pions are expected to be sensitive to a moat regime \cite{Fu:2019hdw, Pisarski:2020gkx}, we propose pion interferometry as a way to detect such a regime in heavy-ion collisions. Based on exploratory results in QCD \cite{Fu:2019hdw}, we expect such a regime in the intermediate density region with $\frac{\mu_B}{T}\gtrsim 4$, where $\mu_B$ is the chemical potential of baryons and $T$ the temperature. This is within reach of present and future heavy-ion experiments at lower beam energies, such as STAR at RHIC, HADES at GSI, CBM at FAIR, NA61/SHINE at CERN \cite{CBM:2016kpk,Bzdak:2019pkr,Almaalol:2022xwv}. 

This paper is organized as follows: We summarize our main results in \Sec{sec:sum}. In \Sec{sec:psh} we develop the formalism to describe particle correlations in a moat regime. This requires techniques of quantum field theory in curved spacetime, which are discussed in \Sec{sec:asy}. We use this to relate $2n$-point functions of quantum fields to $n$-particle spectra in \Sec{sec:spec}, with a special focus on the two-particle spectra relevant for the present purposes. This formalism is then applied to the study of HBT correlations in \Sec{sec:moatHBT}. We first demonstrate in \Sec{sec:HBTG} that the relevant effects are accessible through four-point functions in the Gaussian approximation, and then specify the two-point function of a quasiparticle in a moat regime, which is required as an input. Then,  in \Sec{sec:BW}, we set up a simple model for the energy spectrum and the hypersurface on which the correlations are studied, and numerically analyze the resulting single- and two-particle spectra, as well as the HBT radii in \Sec{sec:res}. We conclude in \Sec{sec:conc}. In a short appendix, \App{app:pai}, we derive the contribution of particle-antiparticle interference to the two-particle spectrum.

\section{Summary of our results}\label{sec:sum}

In order to describe particle spectra in heavy-ion collisions, we first develop a formalism that allows us to compute correlations of particles on curved hypersurfaces in spacetime. This is necessary since, owing to the collision dynamics, fixed thermodynamic conditions, such as a given temperature and density, are only present on hypersurfaces which in general differ from spatial $\mathbb{R}^3$. This is illustrated in \Fig{fig:foliation}. While this idea underlies the well-known Cooper-Frye formalism for single-particle spectra \cite{Cooper:1974mv}, and the hydrodynamic description of two-particle correlations \cite{Makhlin:1986fd}, only spectra of free particles can be described by existing frameworks. In order to study signatures of the QCD phase structure, we need to describe correlations of particles which are subject to in-medium modifications. Based on heuristic arguments, a first attempt to do this for single-particle spectra has been made in Ref.\ \cite{Pisarski:2021qof}, but a microscopic derivation has been lacking. This is provided here. The formalism we develop is based on the quantization of quasiparticles in curved spacetime and their real-time correlation functions. Using a Gaussian approximation, we explicitly derive a relation between the two-particle spectrum and the statistical and spectral function of the particles. This allows us to systematically incorporate in-medium modifications. 

Using this formalism, we show that, in local thermodynamic equilibrium, the in-medium properties of particles contribute to the correlations generated by interference through the spectral functions of the particles evaluated at the average momentum of the particle pair. Hence, the nontrivial information about the phase structure is encoded in the average-momentum dependence of two-particle correlations.
In experimental studies of particle interference it is customary to normalize these correlations,
\begin{align}\label{eq:C}
C(\bold{P},\bold{\Delta P}) = \frac{n_2(\bold{P},\bold{\Delta P})}{n_1\big(\bold{P}+\frac{1}{2}\bold{\Delta P}\big)\, n_1\big(\bold{P}-\frac{1}{2}\bold{\Delta P}\big)}\,,
\end{align}
where $n_1$ and $n_2$ are the single- and two-particle spectra and $\bold{P}$ and $\bold{\Delta P}$ the average and relative momenta of the particle pair.
Our main results are shown in Figs.\ \ref{fig:CratConv} and \ref{fig:CratMoat}. They show ratios of normalized two-particle correlations in three different directions for the relative pair momentum relative to the average momentum, called \emph{out}, \emph{side}, and \emph{long} \cite{Pratt:1984su, Pratt:1986cc}. Figure \ref{fig:CratConv} shows the ratios for particles in the normal phase, i.e.\ where their energy follows \Eq{eq:Eq} with $z=1$ and $w=0$. Figure \ref{fig:CratConv} shows the same ratios, but in a moat regime.
In the moat regime, the ratios show pronounced peaks which are absent in the normal phase.
The location of these peaks is determined by the momentum which minimizes the energy of the particles. This momentum, in turn, is related to the wave number of the underlying spatial modulation. 
If the experimental resolution in the average pair momentum is sufficiently large, these peaks should be experimentally detectable and thus provide a clear signal for a moat regime and, thus, spatially modulated phases in the QCD phase diagram. We note that we use illustrative models to describe the moat regime in this work, so the position of the peak can vary quantitatively depending on
the value of $\bold{p}_{\rm min} \neq 0$, but its existence is nevertheless a robust qualitative prediction of our study. This is the main result of the present work.

\section{Particle spectra on a hypersurface}\label{sec:psh}

We pointed out in Ref.\ \cite{Pisarski:2021qof} that particle spectra in a moat regime cannot be addressed by using the conventional Cooper-Frye formalism \cite{Cooper:1974mv}. This applies to all sorts of particle-number correlations, not only the ones generated by thermodynamic fluctuations studied in Ref.\ \cite{Pisarski:2021qof}. To be most flexible, we therefore aim for a field-theoretic expression of particle spectra in terms of elementary correlations of quantum fields on a (freeze-out) hypersurface $\Sigma$. This is reminiscent of the LSZ reduction formalism.

Particle spectra are related to the phase-space density of the particle number. As such, unless particle number is conserved for the particles under consideration, this notion is only well defined in an `asymptotic' sense. This specifically means that if it is possible to define ladder operators $a_{\bold p}$ and $a_{\bold p}^\dagger$, which annihilate and create a single-particle state with spatial momentum $\bold{p}$ and energy $\omega_{\bold{p}}$, then, e.g., the covariant single- and two-particle spectra are given by
\begin{align}\label{eq:spec}
\begin{split}
n_1(\bold{p}) = (2\pi)^3 \omega_\bold{p} \frac{dN_1}{d^3p} &= \omega_\bold{p} \big\langle a_{\bold p}^\dagger a_{\bold p}  \big\rangle\,,\\
n_2(\bold{p},\bold{q}) = (2\pi)^6 \omega_\bold{p}\omega_\bold{q} \frac{dN_2}{d^3p\,d^3q} &= \omega_\bold{p}\omega_\bold{q} \big\langle a_{\bold p}^\dagger a_{\bold q}^\dagger a_{\bold p}  a_{\bold q}\big\rangle\,,
\end{split}
\end{align}
and analogously for further multi-particle spectra. The $n$-particle spectra for $n\geq 2$ obviously also encode particle-number correlations.
Note that the expectation values do not necessarily have to be vacuum expectation values.

The averages should be with respect to the density matrix on the hypersurface. This can be most conveniently done by quantizing the system on the hypersurface. Another possibility would be to quantize in the rest frame and then boost onto the surface. However, if the system lacks boost invariance, this can be problematic.

One way to compute these spectra is using incoherent sources as discussed, e.g., in Refs.\ \cite{Heinz:1996bs, Weiner:1999th}. Here, we aim at computing these correlations on a hypersurface of the fireball created in heavy-ion collisions and therefore seek to find a formalism more suited to do this directly. To this end, we will develop a new formalism to study spectra on general hypersurfaces.

\subsection{Quasiparticles in foliated spacetime}\label{sec:asy}

As illustrated in \Fig{fig:foliation}, in situations like heavy-ion collisions, a fixed temperature, for example, is typically found on a three-dimensional hypersurface $\Sigma$ of 4$d$ Minkowski spacetime which differs from flat $\mathbb{R}^3$.
Thus, instead of describing these particles in Minkowski space, it is advantageous to describe them in terms of an appropriate \emph{foliation} of spacetime. To this end, we use that any hypersurface $\Sigma$ can be defined by
parametric equations of the form
\begin{align}
x^\mu  = x^\mu(w^i)\,,
\end{align}
where $w^i(x)$ with $i=1,2,3$ are coordinates intrinsic to $\Sigma$ \cite{Poisson:2009pwt}.
We define tangent vectors pointing into `spatial' directions,
\begin{align}
e_i^\mu = \frac{\partial x^\mu}{\partial w^i}\,,
\end{align}
which give rise to the induced metric (first fundamental form) on $\Sigma$,
\begin{align}\label{eq:G}
G_{ij} = -g_{\mu\nu}\, e_i^\mu e_j^\nu\,,
\end{align}
where $g^{\mu\nu}$ is the spacetime metric, i.e., $g^{\mu\nu} = {\rm diag}(1,-1,-1,-1)$ in our case.
Given the coordinates $w^i$, we can express the `temporal' direction via the normal vector
\begin{align}
\hat v^\mu = \frac{e_0^\mu}{\sqrt{e_0^\mu e_{0\mu}}}\,,
\end{align}
with
\begin{align}
e_0^\mu = \bar\epsilon^{\mu\alpha\beta\gamma}\, e_{1\alpha} e_{2\beta} e_{3\gamma}\,,
\end{align}
where $\sqrt{|\det\, g|}\, \bar\epsilon^{\mu\alpha\beta\gamma} = {\rm sign}(\det\, g)\,\epsilon^{\mu\alpha\beta\gamma}$ is the covariant Levi-Civita tensor.
We consider only spacelike hypersurfaces, defined by having a timelike normal vector, $\hat v^\mu \hat v_\mu = +1$.

Using these definitions, we can decompose the ambient metric as
\begin{align}
g^{\mu\nu} = \hat v^\mu \hat v^\nu - \Delta^{\mu\nu}\,,
\end{align}
where $\Delta^{\mu\nu} = G^{ij} e_i^\mu e_j^\nu$ projects onto the hypersurface. This allows us define the temporal and spatial coordinates
\begin{align}\label{eq:coordconv}
x_\parallel = \hat v^\mu  x_\mu\,, \qquad 
\bold{x}_\perp = \bold{e}^\mu  x_\mu\,.
\end{align}
Thus, `time' and `space' are defined as projections of the ambient-space coordinates normal and tangential to $\Sigma$.
This construction defines a foliation of spacetime, where instead of Minkowski space $\{x^0\}\times \mathbb{R}^3$, we describe spacetime as $\{x_\parallel\}\times \Sigma$.

Since the spatial coordinates are the components of tangent vectors $x_\perp^i e_i^\mu$ on $\Sigma$, the corresponding derivatives are covariant derivatives with the induced connection,
\begin{align}
\Gamma_{ijk} = \frac{1}{2}\Bigg( \frac{\partial G_{ij}}{\partial w^k}  + \frac{\partial G_{ik}}{\partial w^j} -\frac{\partial G_{jk}}{\partial w^i}  \Bigg)\,.
\end{align}
This defines the intrinsic covariant derivative, which for a vector field on $\Sigma$, $a_{\perp\, i} = a_\mu e^\mu_i$, is
\begin{align}
\partial_{\perp\,j}\, a_{\perp\, i} = \frac{\partial a_{\perp\,i}}{\partial w^j} - \Gamma^k_{ij}\, a_{\perp\, k}\,.
\end{align}
This can also be expressed in terms of the covariant derivative of the ambient space. Since this is ordinary Minkowski space, this is just the partial derivative, and
\begin{align}
\partial_{\perp\,j}\, a_{\perp\, i} = \partial_\nu\, a_\mu\, e^\mu_i e^\nu_j.
\end{align}
Thus, for a hypersurface embedded into flat space, we can define the intrinsic covariant derivative through the projection
\begin{align}
\partial_{\perp\,i}=e_i^\mu \partial_\mu\,.
\end{align}
In analogy the time derivative is
\begin{align}
\partial_\parallel = \hat v^\nu \partial_\nu\,.
\end{align}
Note that contractions of spatial coordinates always involve the induced metric, e.g., $\bold{a}_\perp\!\!\cdot\bold{b}_\perp = G_{ij}\, a_\perp^i b_\perp^j$. The product of four-vectors becomes $a\!\cdot\!b = a_\parallel b_\parallel - \bold{a}_\perp\!\!\cdot\bold{b}_\perp$.

Spacetime integrals translate into
\begin{align}
d^4x = dx_\parallel\, d^3w\, \bigg| e_0^\mu\frac{\partial x_\mu}{\partial x_\parallel}\bigg| = dx_\parallel\,d^3w\, e_0^\mu\hat v_\mu
\equiv dx_\parallel\, d\Sigma\,, 
\end{align}
and we define
\begin{align}
d\Sigma^\mu = d\Sigma\, \hat v^\mu\,.
\end{align}
With the induced metric defined in \Eq{eq:G}, we can also write $d\Sigma = \sqrt{|\det G|}\, d^3w$.

We assume that the fields we are interested in can be described as quasiparticles on $\Sigma$. The corresponding effective action of a scalar field can be written as
\begin{align}\label{eq:S}
\begin{split}
S = \int\!\!\! dx_\parallel\! \int\!\!\! d\Sigma\bigg\{& \frac{1}{2}\phi(x) \Big[\! -\partial_\parallel^2 + Z\big(-\partial_\perp^2\big)\, \partial_\perp^2 - m^2  \Big] \phi(x)\\
&+ J(x)\phi(x)\bigg\}\,.
\end{split}
\end{align}
We introduced a source $J(x)$ which encodes interactions with external particles or fields.
We allow for a general spatial `wave-function renormalization' $Z\big(-\partial_\perp^2\big)$, which can itself be an arbitrary function of the spatial derivative operator $\partial_\perp^2$. In a free Lorentz-invariant theory $Z=1$. However, if boost invariance is broken, $Z\neq 1$ and higher-order spatial derivative terms can be induced, e.g., by spatial modulations, as in a moat regime. For example, in a low-momentum expansion up to order $\partial_\perp^4$, one recovers the effective action used in Refs.\ \cite{Pisarski:2020dnx,Pisarski:2021qof}.
Since the specific conditions where particles are in a given regime are in general met only on nontrivial hypersurfaces of, e.g., the expanding fireball of a heavy-ion collision, it is most sensible to formulate the resulting effective field theory also on this surface, instead of in flat space.  

Due to the lack of Lorentz invariance, the effective action is frame-dependent. We assume \Eq{eq:S} is the action in the local frame of the particles on $\Sigma$, i.e., in the rest frame of the medium. Because of higher-order spatial derivative terms in the effective action, boosting into a different frame would introduce higher-order time derivatives and the resulting theory will be plagued by Ostrogradsky instabilities. We restrict ourselves to theories with two time derivatives and use ordinary canonical quantization in the following.

Given a specific foliation, we can use the well-known techniques of quantum field theory in curved spacetime to canonically quantize the theory defined by \Eq{eq:S}. To this end, we define the symplectic form
\begin{align}\label{eq:symp}
(\chi_1 , \chi_2 )= i \int\!d\Sigma^\mu \big[ \chi_1^* (\partial_\mu \chi_2) - (\partial_\mu \chi_1^*)\chi_2 \big]\,,
\end{align}
where $\chi_{1,2}(x)$ are solutions of the equations of motion (EoM) for the action \eq{eq:S}. We emphasize that if Lorentz invariance is broken, the symplectic form depends on the choice of the hypersurface $\Sigma$. The free EoM,
\begin{align}\label{eq:freeKGE}
\Big[ \partial_\parallel^2 - Z\big(-\partial_\perp^2\big)\, \partial_\perp^2 + m^2  \Big] \phi_0(x) = 0\,,
\end{align}
are solved by plane waves
\begin{align}\label{eq:u}
u_{\bold{p}_\perp} = \frac{1}{\sqrt{2 \omega_{\bold{p}_\perp}}} e^{-i \bar p \cdot x}\,.
\end{align}
We use a notation where an overline, $\,\overline{\phantom{p}}\,$, indicates that the momentum is on the mass-shell of the particle. Thus $\bar p^\mu$ is the on-shell momentum
and 
\begin{align}\label{eq:disp}
\omega_{\bold{p}_\perp} 
\equiv \hat v^\mu \bar p_\mu 
= \sqrt{Z\big(\bold{p}_\perp^2\big)\, \bold{p}_\perp^2 +m^2}
\end{align}
is the energy of the particle on $\Sigma$.
The waves are normalized such that
\begin{align}\label{eq:unorm}
( u_{\bold{p}_\perp}, u_{\bold{q}_\perp} ) = (2\pi)^3 \delta^{(3)}(\bold{p}_\perp - \bold{q}_\perp) = \int\! d\Sigma\, e^{-i (\bold{p}_\perp-\bold{q}_\perp)\cdot \bold{x}_\perp} \,.
\end{align}
The solution $\phi_0(x)$ of \Eq{eq:freeKGE} can now be expanded in terms of these waves, with the annihilation and creation operators $\tilde a_{\bold{p}_\perp}$ and $\tilde a_{\bold{p}_\perp}^\dagger$ as coefficients,
\begin{align}
\phi_0(x) = \int\!\frac{d^3\bold{p}_\perp}{(2\pi)^3}\,\big( \tilde a_{\bold{p}_\perp} u_{\bold{p}_\perp} + \tilde a_{\bold{p}_\perp}^\dagger u_{\bold{p}_\perp}^* \big)\,.
\end{align}
Using this relation, or through the symplectic form in \Eq{eq:symp}, the ladder operators at time $x_\parallel$ can be expressed as
\begin{align}
\begin{split}
\tilde a_{\bold{p}_\perp} &= \big( u_{\bold{p}_\perp}, \phi_0(x) \big)\\
 &= i \int\!d\Sigma^\mu\, e^{i\bar p \cdot x} \frac{1}{\sqrt{2\omega_{\bold{p}_\perp}}} \big( \partial_\mu - i \bar p_\mu \big) \phi_0(x)\,,\\
\tilde a_{\bold{p}_\perp}^\dagger &= -\big( u_{\bold{p}_\perp}^*, \phi_0(x) \big)\\
 &= -i \int\!d\Sigma^\mu\, e^{-i\bar p \cdot x} \frac{1}{\sqrt{2\omega_{\bold{p}_\perp}}} \big( \partial_\mu + i \bar p_\mu \big) \phi_0(x)\,.
\end{split}
\end{align}
This completes the quantization of the homogeneous case defined by the free EoM in \Eq{eq:freeKGE}. We are, however, interested in the more general case where particles can emerge from  external fields or interactions. The corresponding EoM which follow from \Eq{eq:S} are then
\begin{align}\label{eq:KGE}
\Big[ \partial_\parallel^2 - Z\big(-\partial_\perp^2\big)\, 
\partial_\perp^2 + m^2  \Big] \phi(x) = J(x)\,.
\end{align}
Under the assumption that the source $J$ is only turned on for 
$x_\parallel >0$, and by using the retarded propagator on $\Sigma$,
\begin{align}\label{eq:DR}
D_R(p) = \frac{1}{-(p_\parallel+i\epsilon)^2 + \omega_{\bold{p}_\perp}^2}\,,
\end{align}
\Eq{eq:KGE} is solved by
\begin{align}
\phi(x) = \phi_0(x) + \int\!d^4x'\, D_R(x-x')\, J(x')\,.
\end{align}
Consequently, the shifted ladder operators,
\begin{align}
\begin{split}
a_{\bold{p}_\perp} &= \tilde a_{\bold{p}_\perp} + \frac{i}{\sqrt{2\omega_{\bold{p}_\perp}}} \int\!d^4x\,e^{i\bar p \cdot x}J(x)\,,\\
a_{\bold{p}_\perp}^\dagger &= \tilde a_{\bold{p}_\perp}^\dagger - \frac{i}{\sqrt{2\omega_{\bold{p}_\perp}}} \int\!d^4x\,e^{-i\bar p \cdot x}J^*(x)\,,
\end{split}
\end{align}
are given by
\begin{align}\label{eq:ladder}
\begin{split}
a_{\bold{p}_\perp} &= \big( u_{\bold{p}_\perp}, \phi(x) \big)\\
 &= i \int\!d\Sigma^\mu\, e^{i\bar p \cdot x} \frac{1}{\sqrt{2\omega_{\bold{p}_\perp}}} \big( \partial_\mu - i \bar p_\mu \big) \phi(x)\,,\\
a_{\bold{p}_\perp}^\dagger &= -\big( u_{\bold{p}_\perp}^*, \phi(x) \big)\\
 &= -i \int\!d\Sigma^\mu\, e^{-i\bar p \cdot x} \frac{1}{\sqrt{2\omega_{\bold{p}_\perp}}} \big( \partial_\mu + i \bar p_\mu \big) \phi(x)\,.
\end{split}
\end{align}
We can use these expressions to compute any particle spectrum, such as the ones shown in \Eq{eq:spec}, on an arbitrary spacelike hypersurface $\Sigma$. Furthermore, this equation is valid for arbitrary dispersion relations $\omega_{\bold{p}_\perp}$.

For an ordinary relativistic particle, the relations in \Eq{eq:ladder} are typically the starting point of the derivation of the LSZ reduction formula. There, $x_0 \rightarrow \pm\infty$ in order to define in- and out-states. In contrast, here, $x_\parallel$ can be any time where a quasiparticle picture applies, i.e., where interactions can be considered as classical sources. In the following, we will show that this allows us to express spectra of these particles in terms of their real-time correlation functions. For applications of related ideas to the Schwinger effect and scattering amplitudes, see Refs.\ \cite{Fukushima:2014sia,Bulava:2019kbi}.

\subsection{Spectra from correlators}\label{sec:spec}

With the help of \Eq{eq:ladder} we can express particle spectra on $\Sigma$ in terms of correlation functions of fields. Given the generating functional
\begin{align}\label{eq:Z}
Z[j] = \int\!\mathcal{D}\phi e^{iS[\phi]+i\int\!dx_\parallel\int\!d\Sigma\, j(x)\phi(x)}\,,
\end{align}
we can write in general
\begin{align}\label{eq:gencorr}
\begin{split}
&\big\langle a_{\bold{p}_1}^\dagger \cdots a_{\bold{p}_n}^\dagger a_{\bold{q}_1} \cdots a_{\bold{q}_n} \big\rangle\\ 
&= \lim_{\substack{x_{\parallel,1},\dots,x_{\parallel,n}\rightarrow x_\parallel\\y_{\parallel,1},\dots,y_{\parallel,n}\rightarrow x_\parallel}}\, \Bigg[\prod_{k=1}^n \int\!d\Sigma_{x_k}^\mu\,d\Sigma_{y_k}^{\nu}\, \frac{1}{2\omega_{\bold{p}_{k}}}\, e^{-i\bar p_k \cdot x_k} e^{i \bar q_k \cdot y_k}\\
&\quad\times \big(\partial_{\mu}^{x_k}+i\bar p_{k,\mu}\big)\big(\partial_{\mu}^{y_k}-i\bar q_{k,\mu}\big) \Bigg]\\
&\quad \times \frac{(-i)^{2n}}{Z[0]} \Bigg( \prod_{k=n}^1 \frac{\delta}{\delta j(y_k))}  \Bigg)\Bigg( \prod_{k=n}^1 \frac{\delta}{\delta j(x_k))}  \Bigg)\, Z[j]\Bigg|_{j=0}\,,
\end{split}
\end{align}
Note that the path integral automatically generates time-ordered correlation functions. The above expression only describes spectra on $\Sigma$ after the equal-time limit is taken. This is indicated by the limit in \Eq{eq:gencorr}, where all times $x_{\parallel,k}$ and $y_{\parallel,k}$ are set to the same time $x_\parallel$.
In the following, we explicitly work out the expressions for single-particle spectra. Since this corresponds to two-point functions of the fields, this is sufficient to describe all $n$-particle spectra if the generating functional is Gaussian. All correlations/spectra can then be expressed in terms of two-point functions/single-particle spectra by virtue of Wick's theorem.

For illustration, and since it is most relevant to the application below, see \Sec{sec:HBTG}, we compute the mixed single-particle spectrum,
\begin{align}\label{eq:n1mixed}
n_1(\bold{p}_{\perp},\bold{q}_{\perp}) = \sqrt{\omega_{\bold{p}_\perp} \omega_{\bold{q}_\perp}}\, \big\langle a_{\bold{p}_\perp}^\dagger a_{\bold{q}_\perp}\big\rangle\,.
\end{align}
For $\bold{p}_\perp = \bold{q}_\perp$ 
this reduces to the ordinary single-particle spectrum in \Eq{eq:spec}. 
Using \Eq{eq:ladder} we get
\begin{align}
\begin{split}
n_1(\bold{p}_\perp,\bold{q}_\perp)  &= \frac{1}{2} \lim_{y_\parallel\rightarrow x_\parallel} \int\!d\Sigma_x^\mu\,d\Sigma_y^\nu\,e^{-i\bar p \cdot x}e^{i\bar q \cdot y}\\
&\quad \times \big( \partial^{x}_{\mu}+i\bar p_\mu\big)\big( \partial^{y}_{\nu}-i\bar q_\nu \big) \langle \phi(x)\phi(y) \rangle\,.
\end{split}
\end{align}
Since the particle spectrum is defined at equal time, so is the correlation function on the right-hand side. In order to express this as something resembling a phase-space distribution, we define the average and relative positions
\begin{align}
X = \frac{1}{2}(x+y)\,,\qquad \Delta X = x-y\,. 
\end{align}
By also introducing the average and relative momenta,
\begin{align}\label{eq:momdef}
\begin{split}
P &= \frac{1}{2}(p+q)\,,\qquad \Delta P = p-q\,,\\
\overline P &= \frac{1}{2}(\bar p+\bar q)\,,\qquad \overline{\Delta P} = \bar p-\bar q\,,
\end{split}
\end{align}
the mixed spectrum becomes
\begin{align}\label{eq:ms1}
\begin{split}
n_1(\bold{p}_\perp,\bold{q}_\perp)&= \frac{1}{2} \lim_{\Delta X_\parallel\rightarrow 0} \int\!d\Sigma_X^\mu\,d\Sigma_{\Delta X}^\nu\,e^{-i\overline{\Delta P}\cdot X}e^{-i\overline P \cdot \Delta X}\\
&\quad\times \widetilde\Box_{\mu\nu}\,  \bigg\langle \phi\bigg(X+\frac{1}{2} \Delta X\bigg)\phi\bigg(X-\frac{1}{2}\Delta X\bigg) \bigg\rangle\,.
\end{split}
\end{align}
Here, $d\Sigma_X^\mu = d\Sigma_X \hat v^\mu$ and $d\Sigma_{\Delta X}^\nu= d\Sigma_{\Delta X} \hat v^\nu$ are the integration measures for the average and relative hypersurfaces at times $X_\parallel = \hat v^\mu X_\mu$ and $\Delta X_\parallel = \hat v^\mu \Delta X_\mu$. The operator $\widetilde\Box_{\mu\nu}$ is
\begin{align}
\begin{split}
\widetilde\Box_{\mu\nu} &= \frac{1}{4} \partial_{\mu}^X\partial_{\nu}^X - \frac{1}{2}\partial_{\mu}^X\partial_{\nu}^{\Delta X} + \frac{1}{2} \partial_{\mu}^{\Delta X}\partial_{\nu}^X - \partial_{\mu}^{\Delta X}\partial_{\nu}^{\Delta X}\\
&\quad -\frac{i}{2}\bigg( \overline{P}_\nu - \frac{1}{2} \overline{\Delta P}_\nu \bigg)\big(\partial_{\mu}^X+2\partial_{\mu}^{\Delta X}\big)\\
&\quad+\frac{i}{2}\bigg( \overline{P}_\mu + \frac{1}{2} \overline{\Delta P}_\mu \bigg)\big(\partial_{\nu}^X-2\partial_{\nu}^{\Delta X}\big)\\
&\quad + \bigg( \overline{P}_\mu + \frac{1}{2}\overline{\Delta P}_\mu \bigg) \bigg( \overline{P}_\nu - \frac{1}{2}\overline{\Delta P}_\nu \bigg)\,.
\end{split}
\end{align}
Through the integrations over $d\Sigma_X^\mu$ and $d\Sigma_{\Delta X}^\mu$ this operator is projected normal to these hypersurfaces,
\begin{align}\label{eq:vvbox}
\begin{split}
\hat v^\mu \hat v^\nu \widetilde\Box_{\mu\nu} &= \frac{1}{4}\big(\partial_{\parallel}^X\big)^2 -\big(\partial_{\parallel}^{\Delta X}\big)^2 + \frac{i}{2}\overline{\Delta P}_\parallel \partial_{\parallel}^X - 2 i \overline{P}_\parallel \partial_{\parallel}^{\Delta X}\\
&\quad + \overline{P}_\parallel^2 - \frac{1}{4} \overline{\Delta P}_\parallel^2\,,
\end{split}
\end{align}
where $\overline{P}_\parallel = \frac{1}{2}(\omega_{\bold{p}_\perp}+\omega_{\bold{q}_\perp})$ and $\overline{\Delta P}_\parallel = \omega_{\bold{p}_\perp}-\omega_{\bold{q}_\perp}$.
The integration over $\Delta X$ in \Eq{eq:ms1} is almost a Wigner transformation, which replaces the relative position $\Delta X$ 
by the average momentum $\overline{P}$. What is missing is the $\Delta X_\parallel$ integration. It can be included by introducing and then cancelling the corresponding Fourier transformation (note that this is a transformation with negative momentum $-P$),
\begin{align}
h(\Delta X_\parallel) = \int\!\frac{dP_\parallel}{2\pi}\, e^{iP_\parallel \Delta X_\parallel}\int\!d\Delta X_\parallel^\prime\, e^{-i P_\parallel \Delta X_\parallel^\prime}\,h(\Delta X_\parallel^\prime)\,,
\end{align}
where $h$ is an arbitrary function. Applying this to \Eq{eq:ms1}, we can carry out the full Fourier transform with respect to $\Delta X$. 
We define the Wigner-transformed spectral and statistical functions from the commutators and the anti-commutators of the fields,
\begin{align}\label{eq:rhof}
\begin{split}
\rho(X,P) &=\int\!d\Delta X_\parallel \int\!d\Sigma_{\Delta X}\, e^{iP \cdot \Delta X}\\
&\quad\times \bigg\langle\bigg[ \phi\bigg(X+\frac{1}{2} \Delta X\bigg),\phi\bigg(X-\frac{1}{2}\Delta X\bigg) \bigg]\bigg\rangle\,,
\\
F(X,P) &= \frac{1}{2}\int\!d\Delta X_\parallel \int\!d\Sigma_{\Delta X}\, e^{iP\cdot\Delta X}\\
&\quad\times\bigg\langle\bigg\{ \phi\bigg(X+\frac{1}{2} \Delta X\bigg),\phi\bigg(X-\frac{1}{2} \Delta X\bigg) \bigg\}\bigg\rangle\,.
\end{split}
\end{align}
With this, and taking the limit $\Delta X_\parallel \rightarrow 0$, we finally arrive at
\begin{align}\label{eq:ms}
\begin{split}
n_1(\bold{p}_\perp,\bold{q}_\perp)&= \frac{1}{2}\int\!d\Sigma_X\, e^{-i\overline{\Delta P}\cdot X} \int\! \frac{dP_\parallel}{2\pi}\\
&\quad\times \bigg[ 
\frac{1}{4}\big(\partial_{\parallel}^X\big)^2 + \frac{i}{2}\overline{\Delta P}_\parallel \partial_{\parallel}^X
+\big( P_\parallel + \overline{P}_\parallel \big)^2 - \frac{1}{4} \overline{\Delta P}_\parallel^2
 \bigg]\\
&\quad\times\bigg[ F(X,P) - \frac{1}{2}\rho(X,P) \bigg]\,.
\end{split}
\end{align}
Note that it was important not to go to equal times from the beginning. Otherwise we would not have been able to do the Wigner transformation.

Equation \eqref{eq:ms} is the most general expression we can derive for the mixed particle spectrum. We see that the relative momentum $\overline{\Delta P}$ is correlated with the average location $X$. The Wigner-transformed correlation functions are functions of the average pair momentum $P$.

To get the single-particle spectrum, we set $p = q$, so that $\overline{\Delta P} = 0$ and $P = p$, and
\begin{align}\label{eq:ss}
\begin{split}
n_1(\bold{p}_\perp)&= \frac{1}{2}\int\!d\Sigma_X\,\int\! \frac{dp_\parallel}{2\pi}\, \bigg[ 
\frac{1}{4}\big(\partial_{\parallel}^X\big)^2 +\big( p_\parallel + \omega_{\bold{p}_\perp} \big)^2  \bigg]\\
&\quad\times\bigg[ F(X,p) - \frac{1}{2}\rho(X,p) \bigg]\,,
\end{split}
\end{align}
since the on-shell longitudinal momentum is $\bar p_\parallel = \omega_{\bold{p}_\perp}$.
Although these expressions have been derived assuming a quasiparticle picture, they can in principle be used to define the spectra non-perturbatively and out of equilibrium since they depend on the general spectral and statistical functions of the theory.

We can simplify these expressions by making a few assumptions which are reasonable if one has, as we do, the hydrodynamic regime of a heavy-ion collision in mind. First, we assume that the system is in local thermodynamic equilibrium. This means that $F$ and $\rho$ are related through a generalized fluctuation-dissipation relation (sometimes called the Kadanoff-Baym ansatz) \cite{Kadanoff2018QuantumSM},
\begin{align}
F(X,P) = \bigg[ \frac{1}{2} + f(X,P) \bigg] \rho(X,P)\,,
\end{align}
where $f(X,P)$ is the single-particle distribution. Using this relation, the Boltzmann equation for $f(X,P)$ can be derived from the Kadanoff-Baym equations \cite{Berges:2004yj}. 

Furthermore, we will neglect any derivatives in $X$. This is certainly true for isotropic systems in thermodynamic equilibrium, where only relative positions matter. In general, such gradients are accompanied by additional powers of $\hbar$ and can be neglected in a semi-classical approximation.

Using these approximations, we finally arrive at
\begin{align}\label{eq:s}
\begin{split}
n_1(\bold{p}_\perp,\bold{q}_\perp)&= \frac{1}{2}\int\!d\Sigma_X\, e^{-i\overline{\Delta P}\cdot X} \int\! \frac{dP_\parallel}{2\pi}
\bigg[ 
\big( P_\parallel + \overline{P}_\parallel \big)^2 - \frac{1}{4} \overline{\Delta P}_\parallel^2
 \bigg]\\
&\quad\times f(X,P)\,\rho(X,P)\,,\\
n_1(\bold{p}_\perp) &=\frac{1}{2}\int\!d\Sigma_X\int\! \frac{dp_\parallel}{2\pi}\,
\big( p_\parallel + \omega_{\bold{p}_\perp} \big)^2\, f(X,p)\,\rho(X,p)\,.
\end{split}
\end{align}
These equations can be viewed as generalized Cooper-Frye formulas \cite{Cooper:1974mv}.
In fact, if we use a covariant Bose-Einstein distribution for $f$ and the spectral function of a Lorentz-invariant free particle, $n_1(\bold{p}_\perp)$ reduces to the original expression of Cooper and Frye, see \Sec{sec:quasimoto}. Thus, with our formalism we have derived the phenomenological Cooper-Frye formula from the underlying microscopic quantum field theory. 
Equation (\ref{eq:s}) also explicitly accounts for off-shell effects.
The advantage of our formalism is its generality. The original versions of the Cooper-Frye formalism and HBT, which have been derived in a similar manner as in the present work in Ref.\ \cite{Makhlin:1986fd}, are limited to free particles on the mass-shell. We are not restricted to these limitations here.

\section{HBT on a moat hypersurface}\label{sec:moatHBT}

\subsection{HBT in the Gaussian approximation}\label{sec:HBTG}

With the procedure from above any $n$-particle spectrum can be expressed in terms of $2n$-point functions of fields. For simplicity, we assume that the generating functional in \Eq{eq:Z} is Gaussian. As mentioned above, in the Gaussian approximation $n$-particle spectra can be expressed solely in terms of single-particle ones. 
This turns out to be sufficient for the theoretical description of HBT correlations. 
Here we explicitly work out the two-particle spectrum $n_2(\bold{p}_\perp,\bold{q}_\perp)$, defined in \Eq{eq:spec}. In the Gaussian approximation the 4-point function is
\begin{align}
\begin{split}
&\big\langle \phi(x_1)\phi(x_2)\phi(y_1)\phi(y_2) \big\rangle\\
&\qquad=\big\langle \phi(x_1)\phi(y_1)\big\rangle  \big\langle \phi(x_2)\phi(y_2) \big\rangle\\  
&\qquad\quad+ \big\langle \phi(x_1)\phi(y_2) \big\rangle   \big\langle \phi(x_2)\phi(y_1)\big\rangle\\ 
&\qquad\quad+\big\langle \phi(x_1)\phi(x_2) \big\rangle  \big\langle \phi(y_1)\phi(y_2) \big\rangle\,. 
\end{split}
\end{align}
By comparison with \Eq{eq:gencorr}, we can identify the different contributions:
\begin{align}
\begin{split}
&{\rm (a)}\;\; \big\langle \phi(x_1)\phi(y_1) \big\rangle  \big\langle \phi(x_2)\phi(y_2) \big\rangle\\
&\qquad\longrightarrow \, n_1(\bold{p}_\perp)\,n_1(\bold{q}_\perp)\,,\\[5pt]
&{\rm (b)}\;\; \big\langle \phi(x_1)\phi(y_2) \big\rangle  \big\langle \phi(x_2)\phi(y_1) \big\rangle\\
 &\qquad\longrightarrow \, n_1(\bold{p}_\perp,\bold{q}_\perp)\,n_1(\bold{q}_\perp,\bold{p}_\perp)
  = \big|n_1(\bold{p}_\perp,\bold{q}_\perp)\big|^2\,,\\[5pt]
&{\rm (c)}\;\; \big\langle \phi(x_1)\phi(x_2) \big\rangle  \big\langle \phi(y_1)\phi(y_2) \big\rangle\\
 &\qquad\longrightarrow \omega_{\bold{p}_\perp}\omega_{\bold{q}_\perp}\,\big\langle a_{\bold{p}_\perp}^\dagger a_{\bold{q}_\perp}^\dagger \big\rangle  \big\langle a_{\bold{p}_\perp} a_{\bold{q}_\perp} \big\rangle
 \equiv \big|\bar n_1(\bold{p}_\perp,\bold{q}_\perp)\big|^2\,.
\end{split}
\end{align}
(a) and (b) we already computed. (a) is just a product of two single-particle distributions. (b) stems from particle-particle interference. This is the actual HBT effect. It describes the correlation of identical particles generated by their Bose-Einstein statistics. (c) can be interpreted as particle-antiparticle interference. For particles that are their own antiparticles, like the neutral scalar fields considered here, we cannot distinguish this. Otherwise, one of the ladder operators in each expectation value would correspond to the creation/annihilation operator of an antiparticle.
This is what is called the `surprising effect' in Ref.\ \cite{Weiner:1999th}.

The resulting two-particle spectrum is
\begin{align}\label{eq:c2}
\begin{split}
n_2(\bold{p}_\perp,\bold{q}_\perp)
&= n_1(\bold{p}_\perp)\,n_1(\bold{q}_\perp) + \big|n_1(\bold{p}_\perp,\bold{q}_\perp)\big|^2\\
&\quad+ \big|\bar n_1(\bold{p}_\perp,\bold{q}_\perp)\big|^2\,.
\end{split}
\end{align}
The first two terms are usually considered when studying HBT correlations. Once the spectral function, the single-particle distribution function, and the hypersurface are specified, we can compute this using \Eq{eq:s}. The missing particle-antiparticle interference term is computed in \App{app:pai}. In the following, we set up appropriate models for the required quantities in order to illustrate the relevant qualitative effects.

But first, we note that HBT is often expressed in terms of the emission function (or source function) $S(x,\bold{P}_\perp)$, where $x$ is a spacetime coordinate. It describes the phase-space distribution, i.e., the distribution of spacetime positions and momenta of the particles emitted from a phase-space element $d^3x\,d^3P_\perp$. The particle-particle interference term is then given by the (on-shell) Fourier transformation of the emission function with respect to the relative momentum \cite{Heinz:1996bs,Wiedemann:1999qn},
\begin{align}\label{eq:emf}
n_1(\bold{p}_\perp,\bold{q}_\perp) \approx \int\! d^4x\,e^{-i \overline{\Delta P}\cdot x}\, S(x,\bold{P}_\perp)\,.
\end{align}
In this form, the connection between the behavior of the correlation function and the spacetime structure of the emitting source is most transparent: the range of the correlation in relative momentum $\bold{\Delta P}_\perp$ is related to the inverse size of the source. More precisely, it relates to the region of homogeneity inside the source for a given average momentum $\bold{P}_\perp$ \cite{Makhlin:1987gm}. 

An underlying assumption in \Eq{eq:emf} is that the dependence of the correlation on the relative momentum $\bold{\Delta P}_\perp$ arises solely from the Fourier transformation of the location $x$. As is evident from \Eq{eq:ms}, this is not the case here. Instead, we can define the emission function plus a correction, $S(x,\bold{P}_\perp)+\Delta S(x,\bold{P}_\perp,\bold{\Delta P}_\perp)$, with the emission function
\begin{align}\label{eq:em}
\begin{split}
&S(x,\bold{P}_\perp)\\
&\quad= \frac{1}{2}\int\!d\Sigma_X\, \delta(X-x) \int\! \frac{dP_\parallel}{2\pi}\,
\bigg[ 
\frac{1}{4} \big(\partial^{X}_\parallel\big)^2+\big( P_\parallel + \overline{P}_\parallel \big)^2\bigg]\\
&\quad\quad\times\bigg[ F(X,P) - \frac{1}{2}\rho(X,P) \bigg]\\
&\quad\approx
\frac{1}{2}\int\!d\Sigma_X\, \delta(X-x) \int\! \frac{dP_\parallel}{2\pi}\,
\big( P_\parallel + \overline{P}_\parallel \big)^2 f(X,P)\,\rho(X,P)\,,
\end{split}
\end{align}
and the correction
\begin{align}\label{eq:dem}
\begin{split}
&\Delta S(x,\bold{P}_\perp,\bold{\Delta P}_\perp)\\
&\quad= \frac{1}{4}\int\!d\Sigma_X\, \delta(X-x) \int\! \frac{dP_\parallel}{2\pi}\,
\bigg(i\,\overline{\Delta P}_\parallel \partial^{X}_\parallel
- \frac{1}{2} \overline{\Delta P}_\parallel^2
 \bigg)\\
&\quad\quad\times\bigg[ F(X,P) - \frac{1}{2}\rho(X,P) \bigg]\\
&\quad\approx
-\frac{1}{8}\, \overline{\Delta P}_\parallel^2\int\!d\Sigma_X\, \delta(X-x) \int\! \frac{dP_\parallel}{2\pi}\,
f(X,P)\,\rho(X,P)\,.
\end{split}
\end{align}
In the last lines of both equations we again assumed local thermodynamic equilibrium and neglected gradients in $X$. With this we have
\begin{align}\label{eq:emf2}
n_1(\bold{p}_\perp,\bold{q}_\perp) = \int\! d^4x\,e^{-i \overline{\Delta P}\cdot x}\, \big[S(x,\bold{P}_\perp)+\Delta S(x,\bold{P}_\perp,\bold{\Delta P}_\perp)\big]\,.
\end{align}
Thus, the assumption underlying \Eq{eq:emf} that the HBT correlation can be described as the Fourier transformation of a phase-space distribution is only warranted if the difference of the energies of the on-shell particles on the hypersurface, $\overline{\Delta P}_\parallel$, can be neglected.
This is obviously not true in general. We will discuss the validity of neglecting this contribution in \Sec{sec:res}.

\subsection{Moat quasiparticle}\label{sec:quasimoto}

For the spectral function that enters \Eq{eq:s} we use a similar approximation to the one used in Ref.\ \cite{Pisarski:2021qof}, where thermodynamic fluctuations in a moat regime have been studied. For consistency with our derivation in \Sec{sec:psh}, we assume a quasiparticle
as described by the action in \Eq{eq:S}.
The retarded propagator on $\Sigma$ is then given by \Eq{eq:DR}, which results in the spectral function
\begin{align}\label{eq:quasispec}
\begin{split}
\rho(p) &= 2\, {\rm Im} D_R(p)\\ 
&= \frac{\pi}{\omega_{\bold{p}_\perp}}\,\big[\delta(p_\parallel - \omega_{\bold{p}_\perp})-\delta(p_\parallel + \omega_{\bold{p}_\perp})\big]\,.
\end{split}
\end{align}
In general, any spectral function with a sufficiently sharp peak is suitable for our formalism. We choose a quasiparticle with infinite lifetime for simplicity.

For a quasiparticle with energy $p_\parallel$ on $\Sigma$, the relativistic Boltzmann equation in local thermodynamic equilibrium and in absence of external forces yields the single-particle distribution function
\begin{align}\label{eq:nB}
f(X,p) = \theta(p_\parallel)\,n_B(p_\parallel) = \frac{\theta(p_\parallel)}{e^{p_\parallel/T}-1}\,.
\end{align}
$f$ is a number density in phase space and as such has to be non-negative. This is ensured by the theta function.

The spectral and distribution functions depend on the location of the particle in spacetime through the projection of the momentum onto the hypersurface.
Plugging Eqs.\ \eq{eq:quasispec} and \eq{eq:nB} into Eqs.\ \eq{eq:s} and \eq{eq:barn} yields for the single-particle spectrum,
\begin{align}\label{eq:n1approx}
\begin{split}
n_1(\bold{p}_\perp) =\int\!d\Sigma_X\,\omega_{\bold{p}_\perp}\,n_B(\omega_{\bold{p}_\perp})\,,
\end{split}
\end{align}
for the particle-particle interference contribution,
\begin{align}\label{eq:n1ppapprox}
\begin{split}
&n_1(\bold{p}_\perp,\bold{q}_\perp)=\\
&\quad \frac{1}{4}\int\!d\Sigma_X\Bigg\{ e^{-i\overline{\Delta P}\cdot X} \frac{1}{\omega_{\bold{P}_\perp}}
\bigg[ 
\big( \omega_{\bold{P}_\perp}+ \overline{P}_\parallel \big)^2 - \frac{1}{4} \overline{\Delta P}_\parallel^2
 \bigg]\\
 &\quad\times n_B(\omega_{\bold{P}_\perp})\Bigg\}\,,
\end{split}
\end{align}
and for the particle-antiparticle interference term,
\begin{align}\label{eq:n1paapprox}
\begin{split}
&\bar n_1(\bold{p}_\perp,\bold{q}_\perp) =\\
&\quad-\frac{1}{8}\int\!d\Sigma_X\Bigg\{ e^{-2i\overline{P} \cdot X}\frac{1}{\omega_{\bold{\Delta P}_\perp/2}}\bigg[\big(2\omega_{\bold{\Delta P}_\perp/2} + \overline{\Delta P}_\parallel  \big)^2\\
&\quad- 4 \overline{P}_\parallel^2 \bigg]\, n_B(\omega_{\bold{\Delta P}_\perp/2})\Bigg\}\,.
\end{split}
\end{align}
We see that with the present approximations, we recover the original Cooper-Frye formula for the single-particle spectrum $n_1(\bold{p}_\perp)$, but with a more general dispersion relation. We have therefore provided a microscopic derivation for the expression proposed in Ref.\ \cite{Pisarski:2021qof}.

The interference terms (\ref{eq:n1ppapprox}) and (\ref{eq:n1paapprox}) receive contributions from both on- and off-shell momenta of the individual particles. In the quasiparticle picture also the on-shell average and relative energies $\omega_{\bold{P}_\perp}$ and $\omega_{\bold{\Delta P}_\perp/2}$ contribute. They are manifestly different from the average and relative energies of the on-shell particles, $\overline{P}_\parallel$ and $\overline{\Delta P}_\parallel$. This reflects the generality of our formalism, as it is able to capture off-shell effects by taking into account the full spectrum of the particles involved. In previous work it was necessary to resort to on-shell approximations to describe interference \cite{Heinz:1996bs}.

Furthermore, our microscopic derivation reveals a different momentum dependence of the  particle-particle interference contribution \eq{eq:n1ppapprox} as compared to the literature, cf., e.g., Refs.\ \cite{Schlei:1992jj, Heinz:1996bs}. In previous work where correlations on a sharp hypersurface have been considered (as is customary in hydrodynamical applications), it has always been assumed that interference can be described by the ordinary Cooper-Frye formula for average momenta, and an additional Fourier transformation with respect to the average position of the particles \footnote{See, e.g., Eq.\ (85) in \cite{Heinz:1996bs}}. Our results show that this is not correct, as the particles on the hypersurface cannot be described by their average momentum alone. 

To be specific, in addition to the assumptions that lead from the most general Eqs.\ \eq{eq:ms}, \eq{eq:ss}, and \eq{eq:as} to the above equations, our results reduce to the ones in the literature \cite{Heinz:1996bs,Wiedemann:1999qn} using the following assumptions: First, the contribution from particle-antiparticle interference, $\bar n_1(\bold{p}_\perp,\bold{q}_\perp)$ needs to be neglected. Second, the difference between the on-shell average energy $\omega_{\bold{P}_\perp}$ and the average of the on-shell energies $\overline{P}_\parallel$ has to be neglected, $\overline{P}_\parallel = \omega_{\bold{P}_\perp}$. Furthermore, the relative on-shell energy needs to be neglected as well, $\overline{\Delta P}_\parallel = 0$. Evidently, these assumptions cannot be true in general, so the known equations in the literature have to be considered approximations of the more general ones derived here. Since a quantitative analysis is not in the scope of this work, we defer an analysis of the validity of these approximations to future work.

\begin{figure}[t]
\centering
\includegraphics[width=.9\columnwidth]{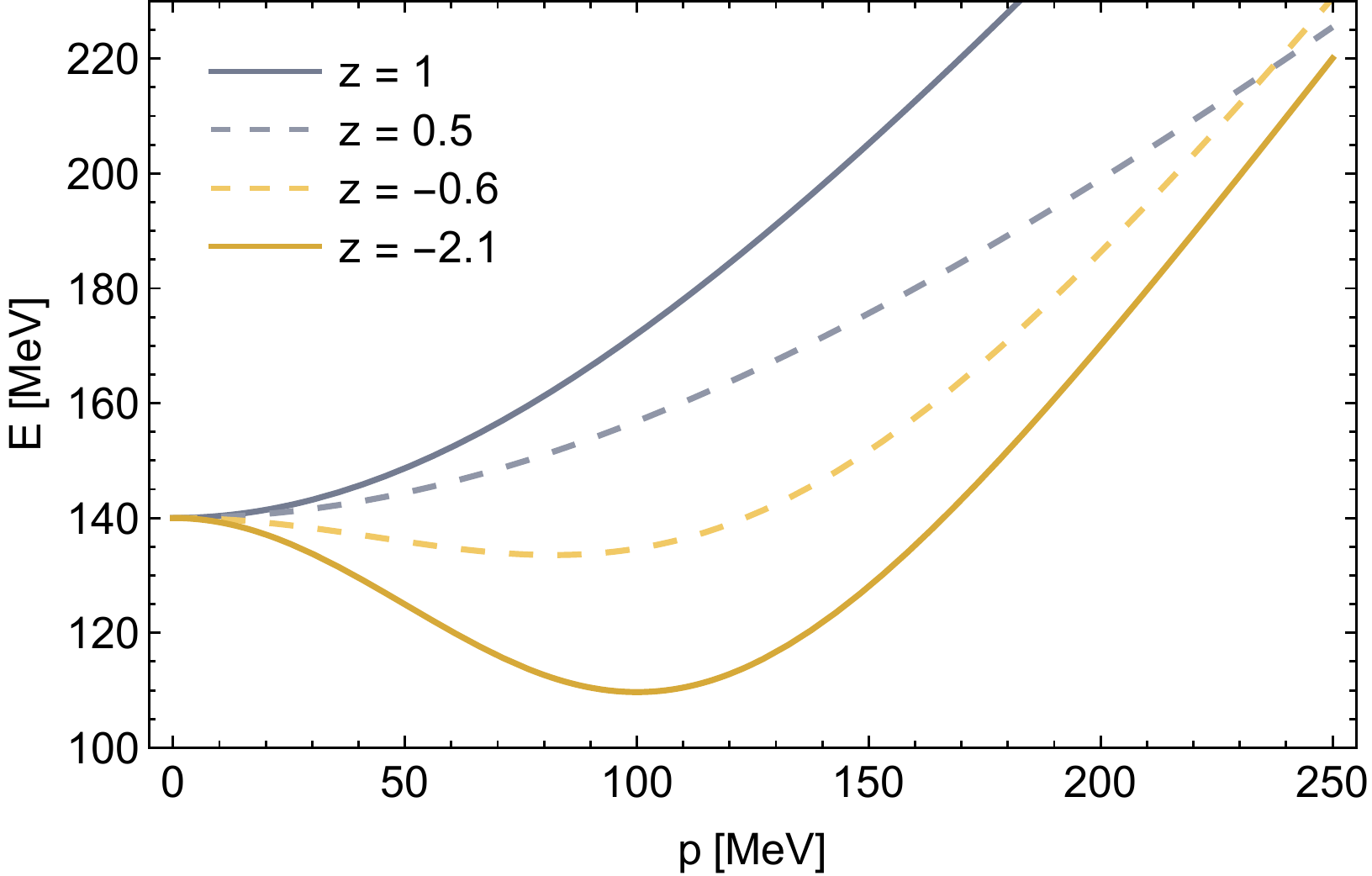}
\caption{Energy as a function of momentum. The gray lines are in the normal phase and the yellow lines are in a moat regime. $z$ is the coefficient of the momentum-squared term of the energy. In the normal phase it is $E_{\bold{p}} = \sqrt{z\, \bold{p^2}+m^2}$ with $z>0$. In the moat regime, the energy is defined in \Eq{eq:Zansatz}, where $z=1-\frac{\lambda^2}{M^2}$. Negative $z$ signals a moat regime.}
\label{fig:disp}
\end{figure}

\subsection{A simple model}\label{sec:BW}

We want to apply the formalism developed here to particles in a moat regime. To this end, we need to specify the wave-function renormalization $Z\big(\bold{p}_\perp^2\big)$, or, equivalently, the dispersion relation \eq{eq:disp}. For simplicity, we specify $Z$ in ambient-space coordinates and then project the resulting on-shell momentum normal to $\Sigma$.
Our ansatz is based on two requirements. First, the resulting energy $\omega_{\bold{p}_\perp}$ should have a minimum at nonzero momentum and, second, it should have a well-defined large-momentum limit. The latter requirement entails that the energy grows linearly with momentum for large momenta, as otherwise causality would be violated. Motivated by Ref.\ \cite{Schindler:2019ugo}, we choose
\begin{align}\label{eq:Zansatz}
\begin{split}
E_{\bold{p}}^2&=  Z\big(\bold{p}^2\big)\,\bold{p}^2+m^2\\
&= \bigg(1 - \frac{\lambda^2}{\bold{p}^2+M^2}\bigg)\,\bold{p}^2+m^2\\
&\approx \bigg(1 - \frac{\lambda^2}{M^2}\bigg)\,\bold{p}^2 +\frac{\lambda^2}{M^4}\bold{p}^4 + \mathcal{O}\big(\bold{p}^6\big)\,,
\end{split}
\end{align}
where in the second line we did a low-momentum expansion. Thus, for small momenta we have the energy $E_{\bold{p}} \approx \sqrt{z\, \bold{p} + w\, \bold{p}^4 + m^2}$ mentioned in \Eq{eq:Eq}, where $z= 1-\frac{\lambda^2}{M^2}$ and $w = \frac{\lambda^2}{M^4}$. Such a dispersion relation has been assumed in Refs.\ \cite{Pisarski:2020dnx,Pisarski:2021qof}.
The energy has a negative $\bold{p}^2$-coefficient if $\frac{\lambda^2}{M^2} >0$, and the system is in a moat regime. 
In this case, the energy is minimized at a nonzero momentum, $|\bold{p}_{\rm min}| = \sqrt{M(\lambda-M)}$, with minimal energy $E_{\rm min} = \sqrt{m^2-(\lambda-M)^2}$. This is illustrated in \Fig{fig:disp}.

Equation (\ref{eq:Zansatz}) yields the on-shell momentum $\bar p^\mu = (E_\bold{p},\bold{p})$, and the resulting energy on the hypersurface is 
$\omega_{\bold{p}_\perp} = \hat v^\mu \bar p_\mu$. The advantage of our ansatz for $Z$ in \Eq{eq:Zansatz} is that the corresponding particles still behave as free relativistic ones at large momenta, $\bold{p}^2 \gtrsim M^2$.

It is left to specify the hypersurface $\Sigma_X$. 
In general it can be extracted from hydrodynamical simulations. To illustrate the basic physics we use a simple model.

We start by setting up the coordinates we use and the parametrization of $\Sigma_X$.
We assume that the hypersurface $\Sigma_X$ is defined by a constant proper time, $\tau = \tau_\Sigma$, in `beam-direction' $X_3$, where $\tau = \sqrt{X_0^2-X_3^2}$ and $X_\mu$ are the Cartesian coordinates of the ambient Minkowski space. This is a common and phenomenologically successful assumption in the context of heavy-ion collisions \cite{Florkowski:2010zz}. 
For the coordinates on $\Sigma_X$ we can then choose the longitudinal spacetime rapidity $\eta_\parallel = {\rm arctanh}\,(X^3/X^0)$, transverse radius $r = \sqrt{X_1^2+X_2^2}$, and the azimuthal angle $\phi = \arctan (X^2/X^1)$.
The resulting normal vector is
\begin{align}
\hat v^\mu = \begin{pmatrix} \cosh\eta_\parallel \\ 0 \\ 0 \\ \sinh\eta_\parallel\end{pmatrix}\,.
\end{align}
The induced metric and its inverse are for $w_1 =\eta_\parallel,\,w_2=r,\,w_3=\phi$
\begin{align}
G_{ij} = \begin{pmatrix} \tau_f^2 & 0& 0 \\ 0 & 1 & 0 \\ 0 & 0 & r^2  \end{pmatrix}\,,\qquad
G^{ij} = \begin{pmatrix} \tau_f^{-2}& 0& 0 \\ 0 & 1 & 0 \\ 0 & 0 & r^{-2}  \end{pmatrix}\,,
\end{align}
and the projection operator onto $\Sigma_X$ is
\begin{align}
\Delta^{\mu\nu} = \begin{pmatrix} 
\sinh^2 \eta_\parallel & 0 & 0 & \sinh \eta_\parallel\cosh \eta_\parallel\\
0 & 1 & 0 & 0\\
0 & 0 & 1 & 0\\
\sinh \eta_\parallel\cosh \eta_\parallel & 0 & 0 & \cosh^2 \eta_\parallel
 \end{pmatrix}\,.
\end{align}
With this we get
\begin{align}
X_\parallel = \tau_f\,,\qquad
\bold{X}_\perp =\begin{pmatrix} 0 \\ -r \\ 0 \end{pmatrix}\,.
\end{align}
We additionally assume that the system has a maximal radial extent $\bar R$. The resulting integration measure is
\begin{align}
d\Sigma_X = \tau_\Sigma r\theta(\bar R-r)\, d\eta_\parallel\,dr\, d\phi\,. 
\end{align}
Motivated by blast-wave model fits to experimental data at small beam energy \cite{Zhang:2016tbf}, we choose $\bar R = 8$\,fm and $\tau_\Sigma = 5\, {\rm fm/c}$.

For the parametrization of the momenta, we take into account that the system only possesses rotational invariance in transverse direction. Thus, we use energy $p_0$, momentum in beam direction $p_3$, transverse momentum $p_T = \sqrt{p_1^2+p_2^2}$ and azimuthal angle $\phi_p = {\rm arctan}(p^2/p^1)$, which lead to
\begin{align}\label{eq:mompar}
\begin{split}
p_\parallel &= p_0 \cosh\eta_\parallel -p_3 \sinh\eta_\parallel\,,\\
\bold{p}_\perp &= 
\begin{pmatrix}
p_0 \tau_f \sinh\eta_\parallel-p_3 \tau_f \cosh\eta_\parallel\\
-p_T \cos(\phi_p -\phi)\\
-p_T r \sin(\phi_p -\phi)
\end{pmatrix}\,.
\end{split}
\end{align}

\subsection{Numerical results}\label{sec:res}

Two-particle spectra are in general six-dimensional functions of the spatial momenta of the two particles. We express these in terms of the coordinates of the ambient Minkowski space. Since this is customary in experiments, we choose the center-of-momentum frame of the particle pair in beam direction, $P_3 = 0$. By fixing the remaining components of the average momentum $P$, we can define the correlation in \emph{out}, \emph{side}, and \emph{long} direction. To this end, we choose $\phi_P = 0$, so that the average momentum points in 1-direction, resulting in
\begin{align}\label{eq:Pdef}
\bold{P} = \begin{pmatrix} P_T\\ 0\\ 0 \end{pmatrix}\,.
\end{align}
We then define the relative momenta
\begin{align}\label{eq:DPdef}
\bold{\Delta P}_{\rm out} = \begin{pmatrix} \Delta P_T \\ 0 \\ 0 \end{pmatrix}\,,\;
\bold{\Delta P}_{\rm side} = \begin{pmatrix} 0 \\ \Delta P_T \\ 0 \end{pmatrix}\,,\;
\bold{\Delta P}_{\rm long} = \begin{pmatrix} 0 \\ 0 \\ \Delta P_3 \end{pmatrix}\,.
\end{align}
In the following, we express the spectra in terms of these momenta, e.g., $n_1(\bold{p}_\perp,\bold{q}_\perp) = n_1\bigg(\bold{P}_\perp+\frac{1}{2}\bold{\Delta P}_\perp,\,\bold{P}_\perp-\frac{1}{2}\bold{\Delta P}_\perp\bigg) \equiv n_1(\bold{P},\bold{\Delta P})$.
Since our parametrization is in terms of the momenta in ambient-space coordinates, we also express the correlations in terms of these in the following.

With a specific parametrization of the momenta, we can address the validity of assuming that the emission function discussed in \Sec{sec:HBTG} only depends on the average momentum. From Eqs.\ \eq{eq:Pdef} and \eq{eq:DPdef} follows that the particle momenta $\bold{P}\pm \bold{\Delta P}$ have equal magnitude in side- and long-direction. This leads to vanishing relative on-shell energies in ambient-space coordinates, $\overline{\Delta P}_0$. However, the relative energy on the hypersurface, $\overline{\Delta P}_\parallel = \overline{\Delta P}_0 \cosh \eta_\parallel - \Delta P_3 \sinh\eta_\parallel$, is relevant for the correlations. This only leads to a non-vanishing contribution if both $\overline{\Delta P}_0$ and $\Delta P_3$ vanish. In our case, this is true for the side-correlation. Hence, only the side-correlation can be described by an emission function $S(x,\bold{P}_\perp)$ in \Eq{eq:em}, the emission functions in out and long directions also depend on the correction $\Delta S(x,\bold{P}_\perp,\bold{\Delta P}_\perp)$ in \Eq{eq:dem}. 

\begin{figure}[t]
\centering
\includegraphics[width=.9\columnwidth]{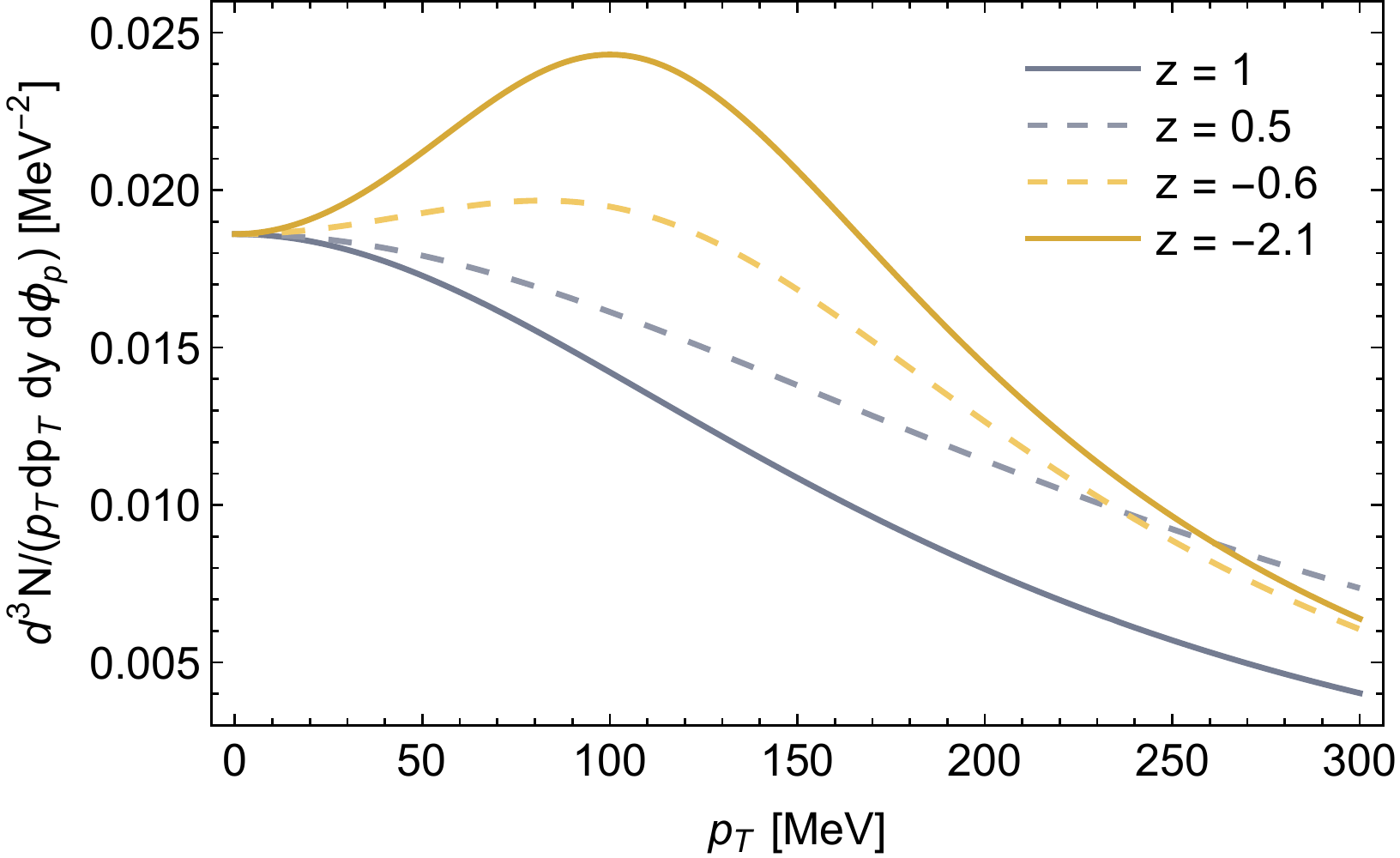}
\caption{Single-particle spectrum from \Eq{eq:n1approx}. The parameters are the same as in \Fig{fig:disp}.}
\label{fig:spec}
\end{figure}

To fix the remaining free parameters, we choose $m=140$\,MeV for the effective mass since it has been predicted in 
Ref.\ \cite{Fu:2019hdw} that pions are sensitive to the moat regime, and are also commonly used to perform experimental HBT studies. We assume that the hypersurface $\Sigma_X$ is defined by a fixed temperature of $T = 115$\,MeV, which is consistent with the anticipated location of a moat regime in the QCD phase diagram \cite{Fu:2019hdw, Rennecke:2021ovl}. Unless otherwise stated, $\lambda=202\,{\rm MeV}$ and $M=115\,{\rm MeV}$ for the parameters of the dispersion relation in \Eq{eq:Zansatz}. Since $\frac{\lambda^2}{M^2}>1$ in this case, the system is in a moat regime for these parameters. This choice leads to $|\bold{p}_{\rm min}| = 100\,{\rm MeV}$ and $E_{\rm min} = 110\,{\rm MeV}$.

\begin{figure*}[t]
\centering
\includegraphics[width=.33\textwidth]{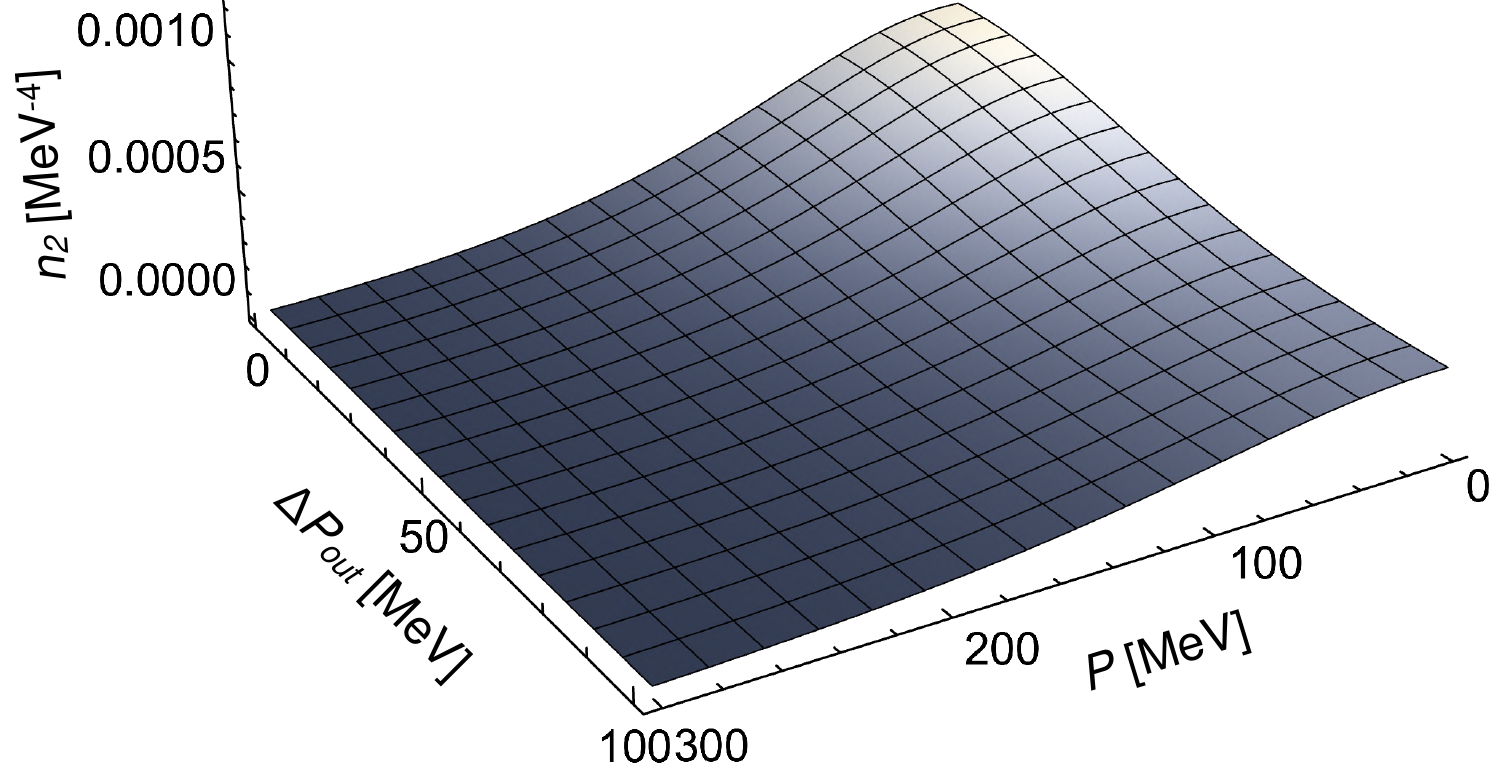}
\hfill
\includegraphics[width=.33\textwidth]{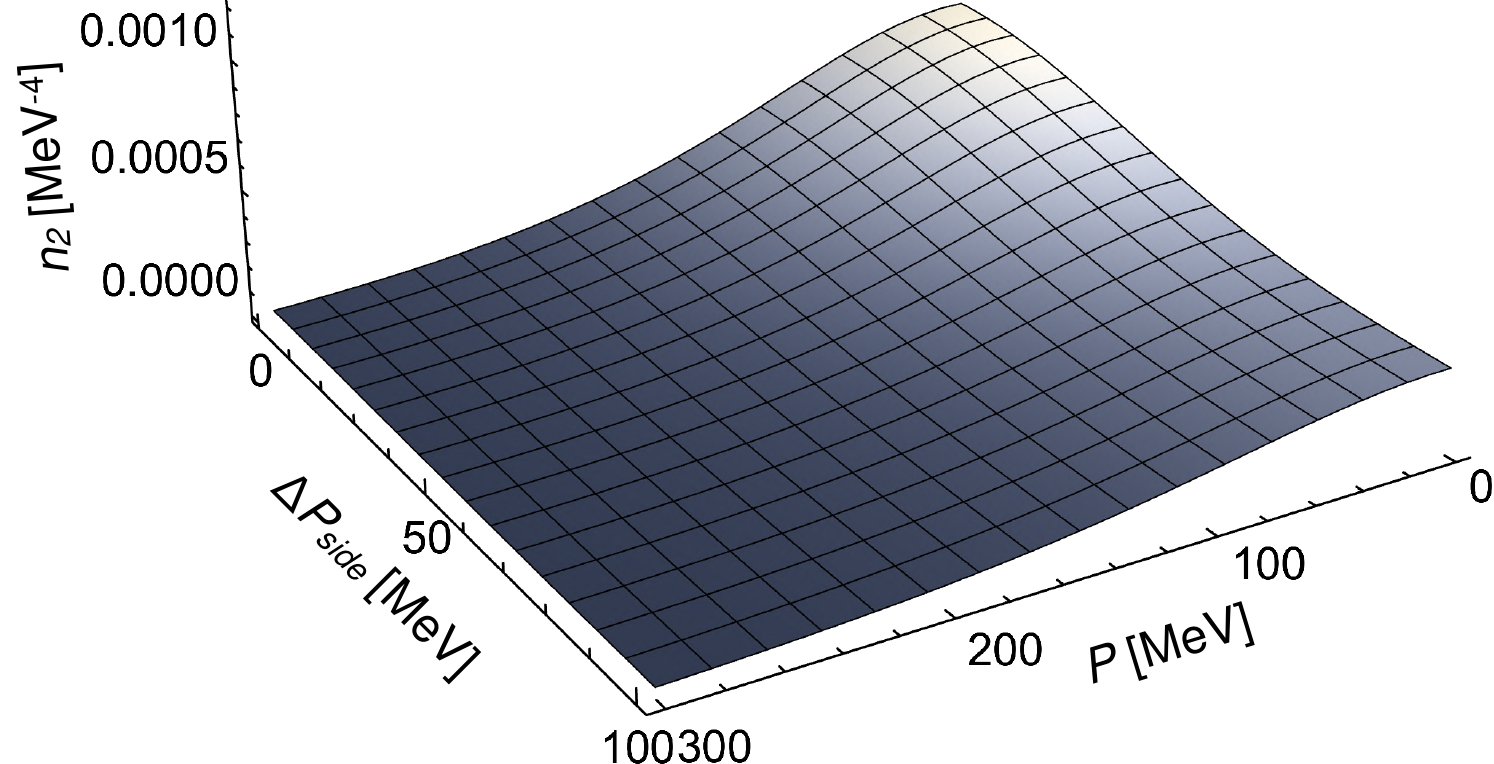}
\hfill
\includegraphics[width=.33\textwidth]{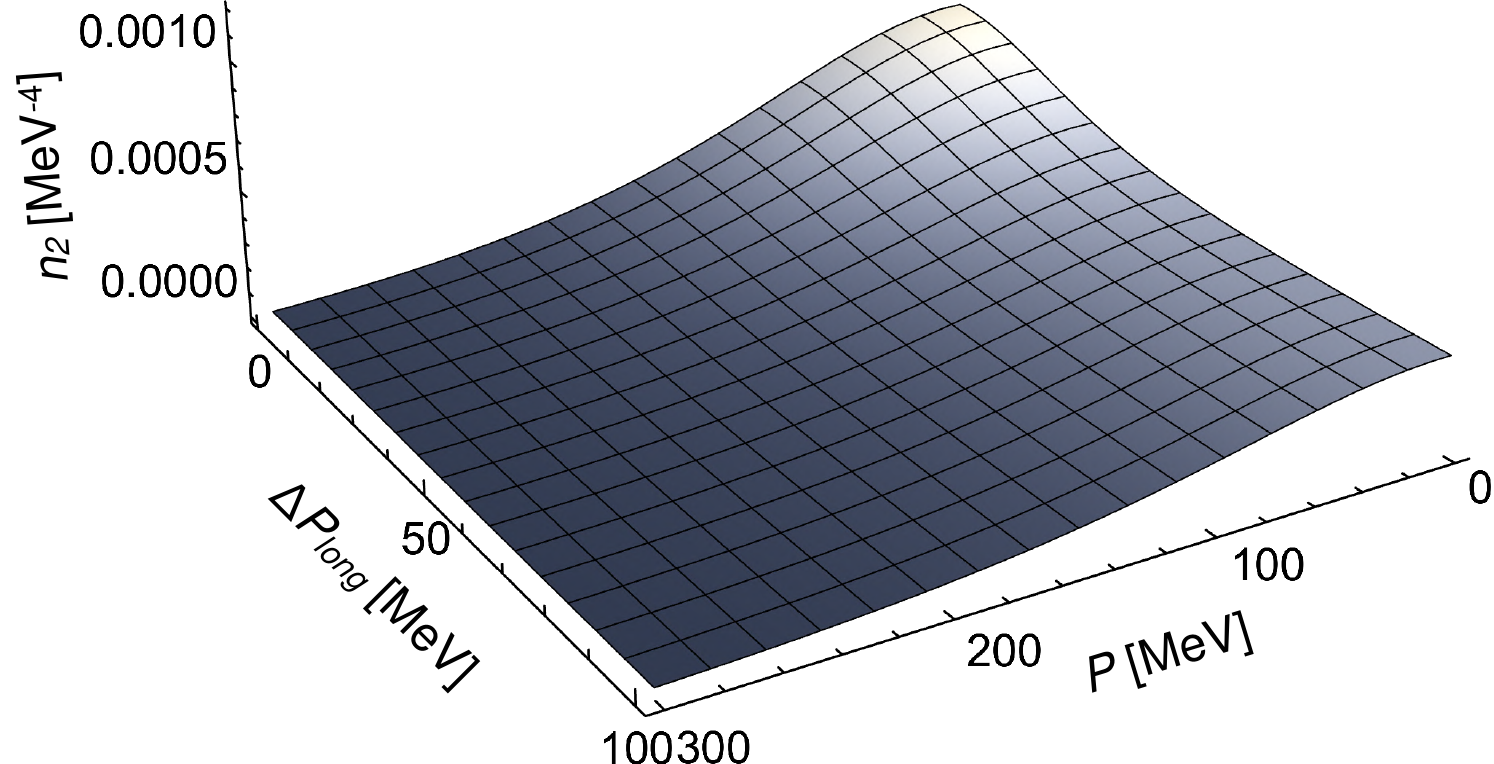}
\caption{Two-particle spectrum $n_2(\bold{P},\bold{\Delta P})$ from \Eq{eq:n1ppapprox} in out, side, and long direction in the normal phase. We denote $P = |\bold{P}|$ and $\Delta P = |\bold{\Delta P}|$.}
\label{fig:n2Conv}
\end{figure*}
\begin{figure*}[t]
\centering
\includegraphics[width=.33\textwidth]{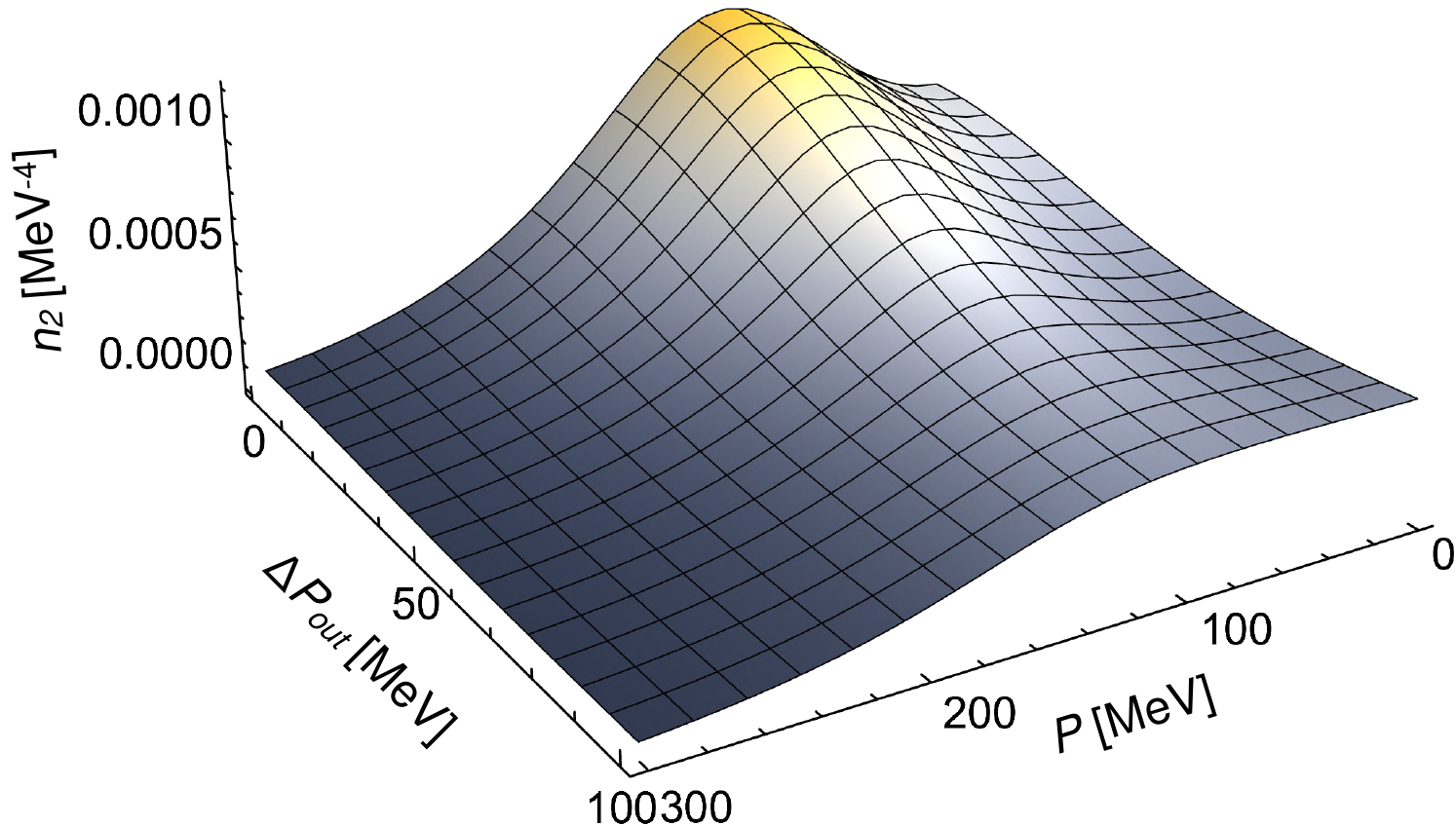}
\hfill
\includegraphics[width=.33\textwidth]{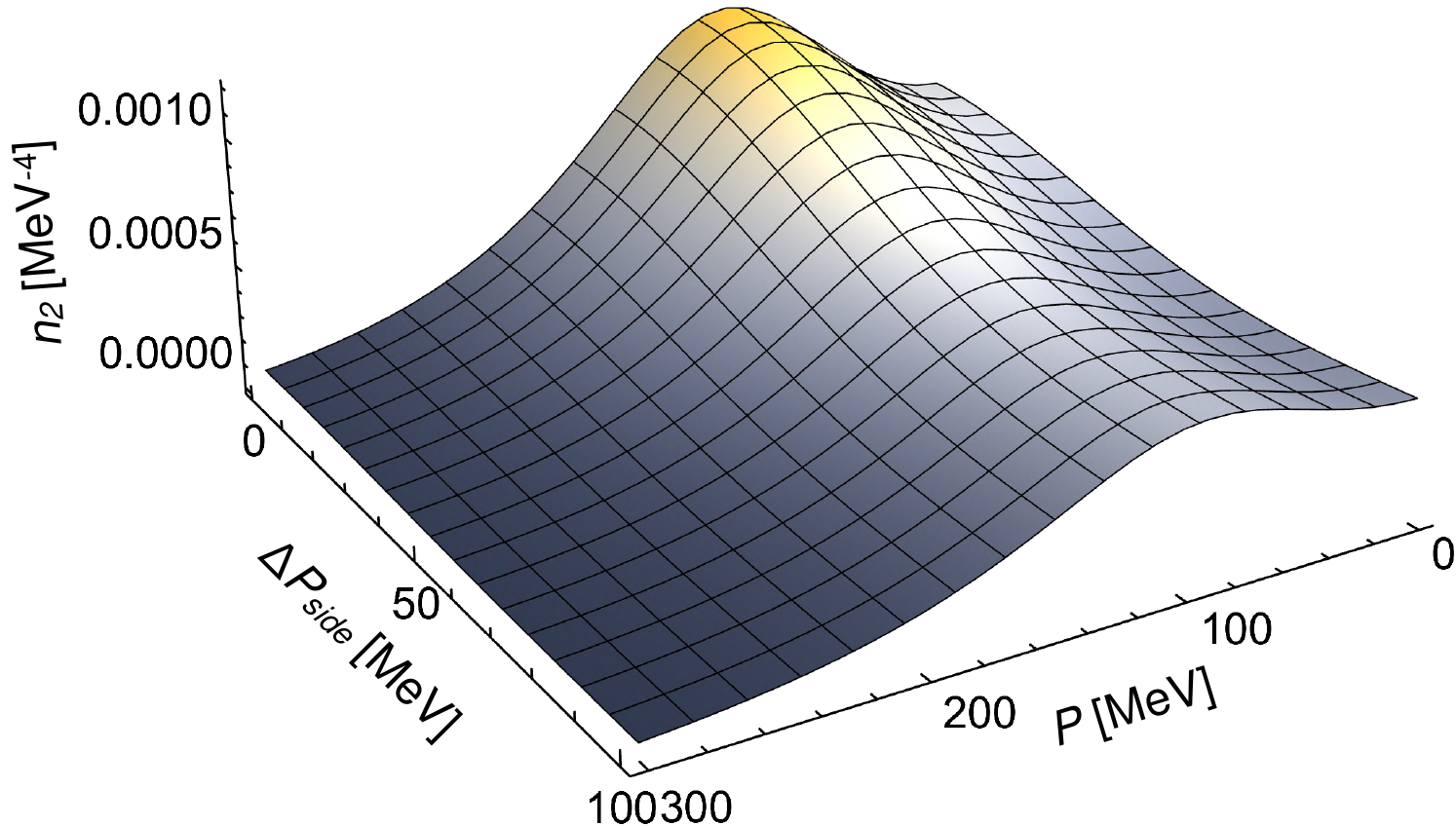}
\hfill
\includegraphics[width=.33\textwidth]{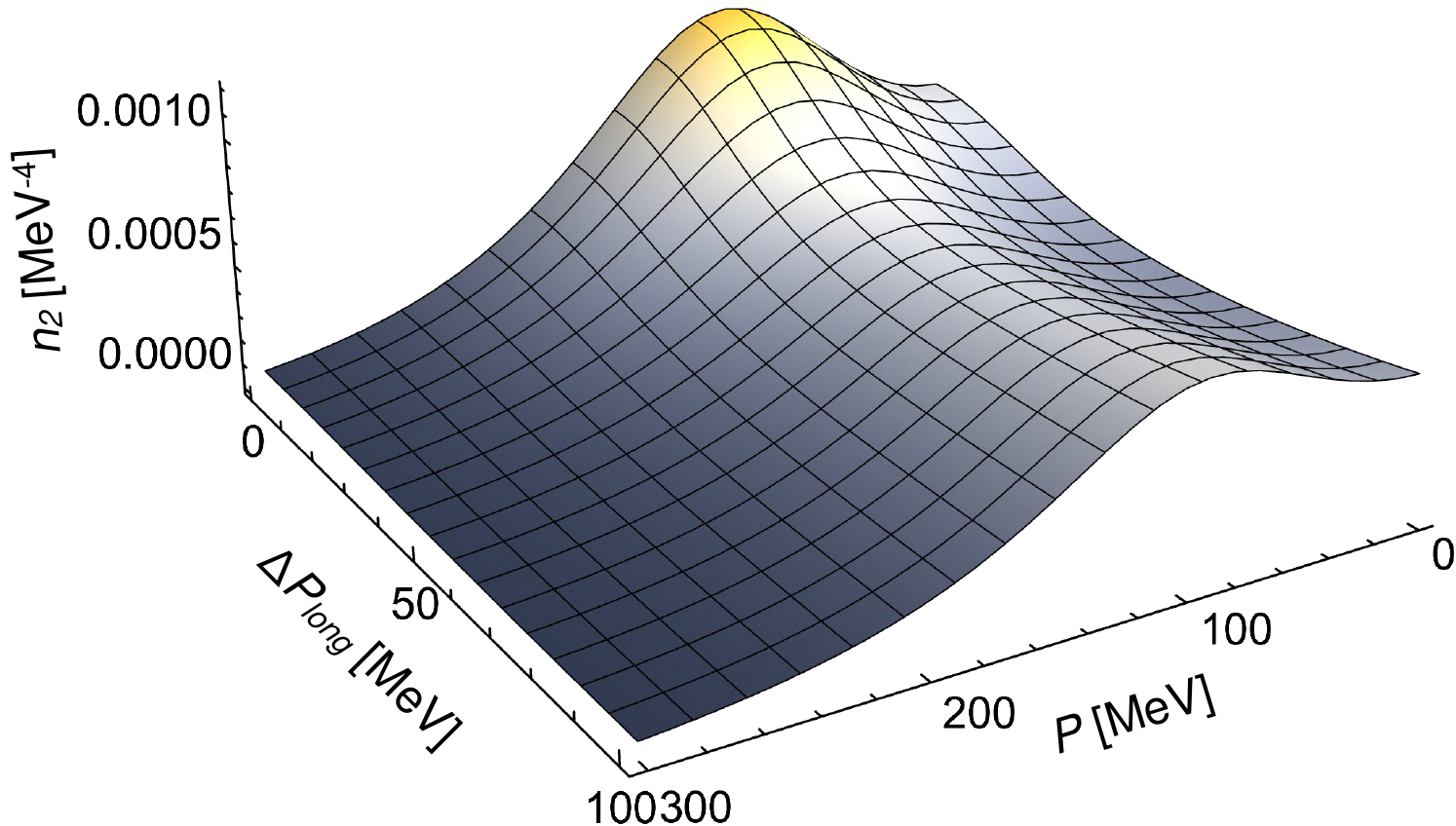}
\caption{Two-particle spectrum $n_2(\bold{P},\bold{\Delta P})$ from \Eq{eq:n1ppapprox} in out, side, and long direction in the moat regime. We denote $P = |\bold{P}|$ and $\Delta P = |\bold{\Delta P}|$.}
\label{fig:n2Moat}
\end{figure*}

In the following, we will always compare results in the normal phase to the moat regime. This is distinguished by the underlying dispersion relation,
\begin{align}
\begin{split}
{\rm \emph{normal phase}}:&\quad \omega_{\bold{p}_\perp}=\sqrt{\bold{p}_\perp^2+m^2}\\
{\rm \emph{moat regime}}:&\quad \omega_{\bold{p}_\perp}=\sqrt{Z(\bold{p}_\perp^2)\,\bold{p}_\perp^2+m^2}
\end{split}
\end{align}
with $Z(\bold{p}_\perp^2)$ specified at the beginning of \Sec{sec:BW} and the parameters given above.

\subsubsection{Single-particle spectrum}

The single-particle spectrum $n_1(\bold{p})$ on $\Sigma_X$ as a function of transverse momentum is shown in \Fig{fig:spec} at $p_3 = 0$ and $\phi_p=0$. The grey lines are for a particle in the normal phase and the yellow lines are in a moat regime. As also shown in Ref.\ \cite{Pisarski:2021qof}, the spectrum of a particle in a moat regime shows a characteristic non-monotonic $p_T$-dependence. The initial increase of the spectrum with increasing momentum in the moat regime results from the decrease in energy in a moat regime. The location of the resulting peak is related to the wave number of the underlying spatial modulation, determined by the momentum at the bottom of the moat, $\bold{p}_{\rm min}$.

\begin{figure*}[t]
\centering
\includegraphics[width=.305\textwidth]{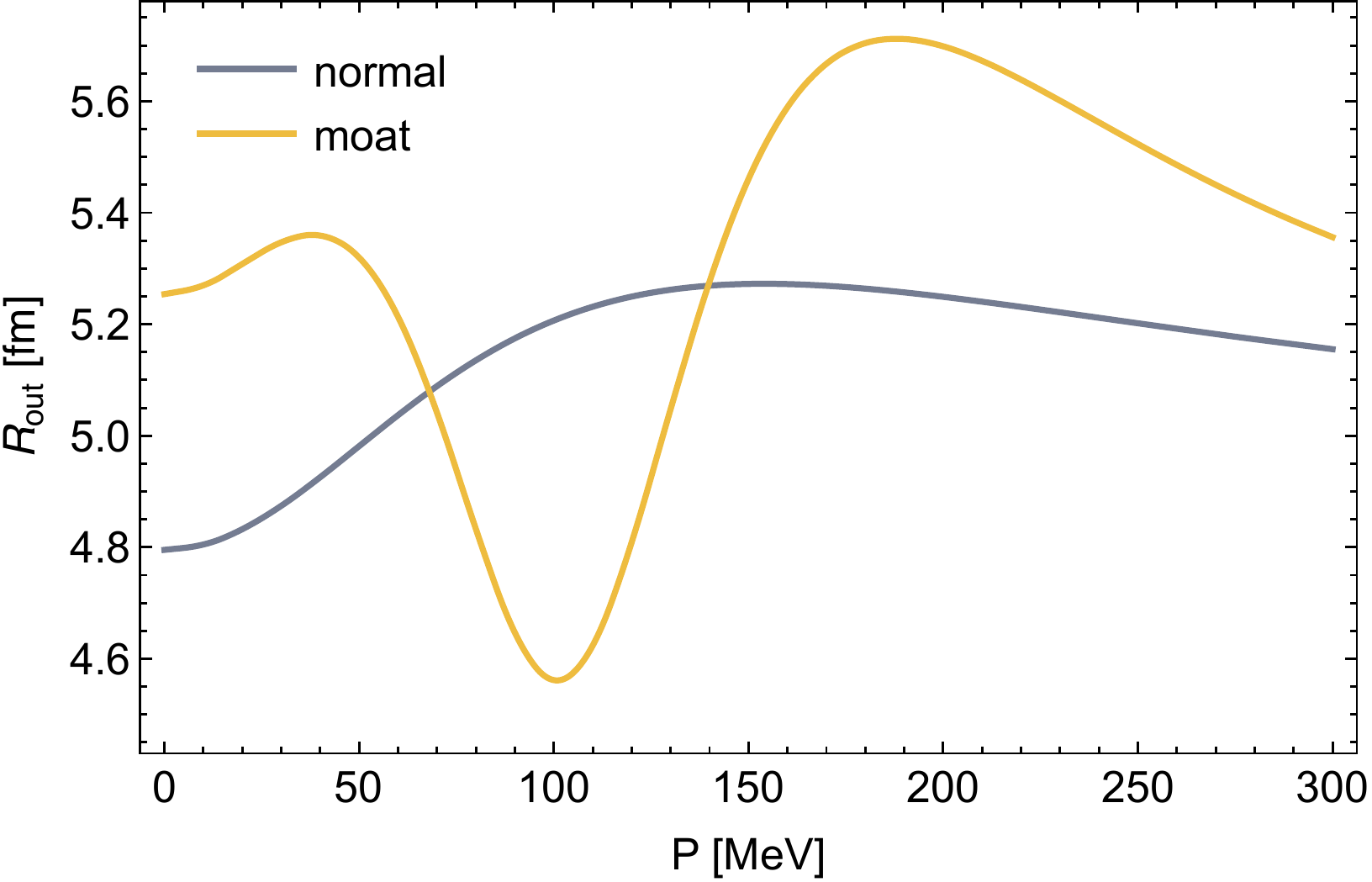}
\hfill
\includegraphics[width=.31\textwidth]{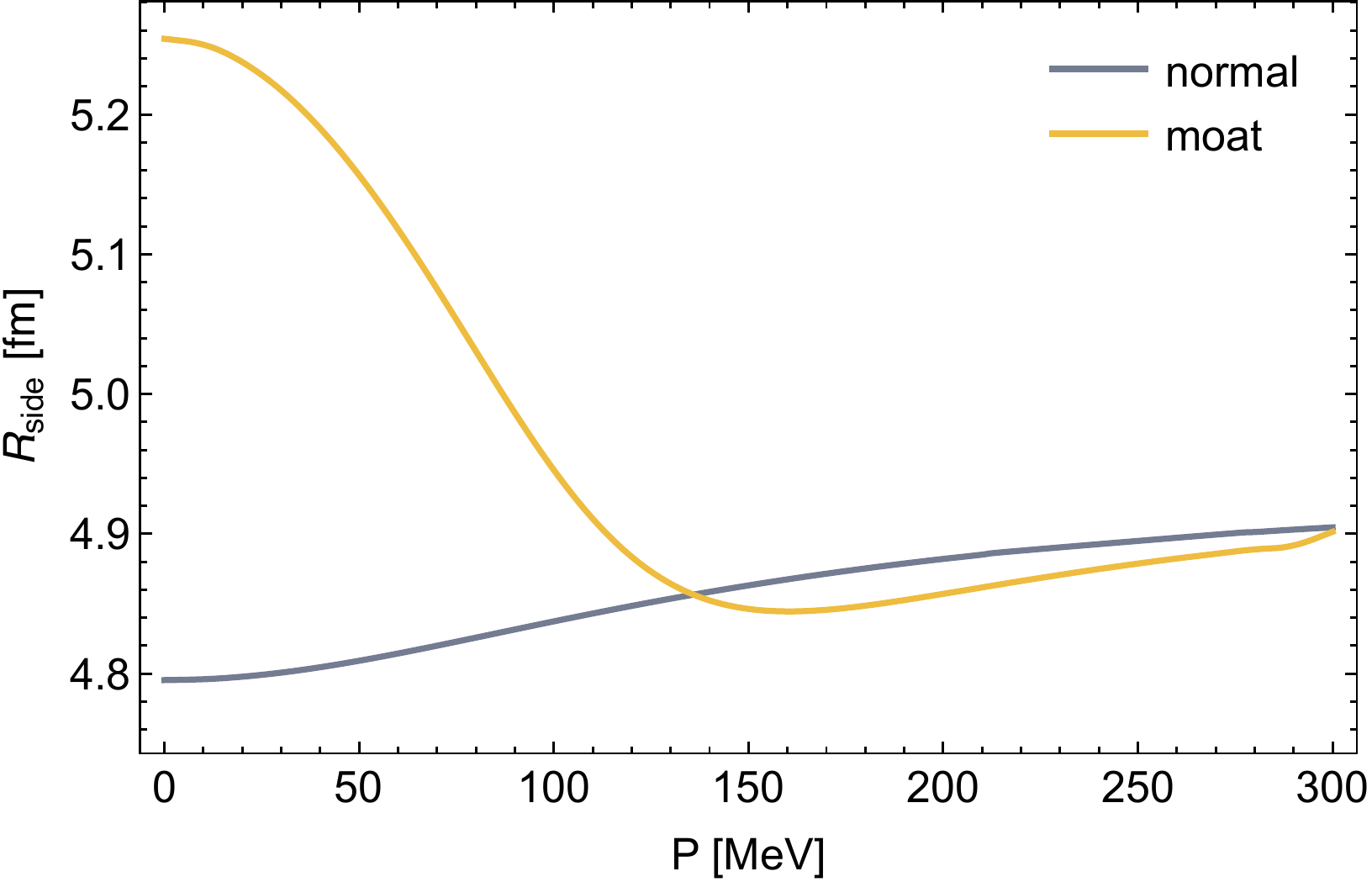}
\hfill
\includegraphics[width=.30\textwidth]{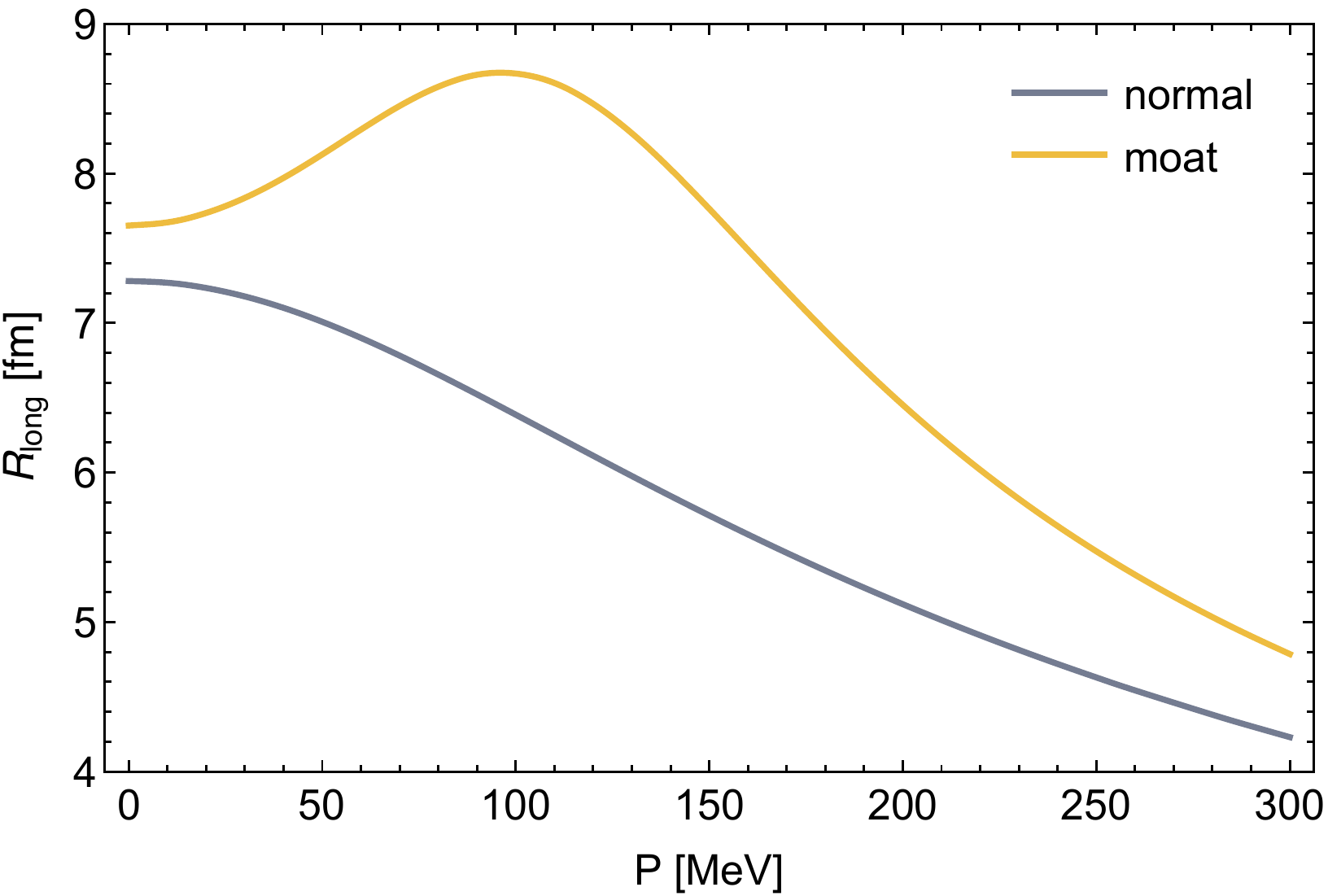}
\caption{HBT radii $R_{\rm out}$, $R_{\rm side}$, and $R_{\rm long}$ as functions of the average momentum in the normal phase (gray) and the moat regime (yellow).}
\label{fig:RHBT}
\end{figure*}

\subsubsection{Two-particle spectrum}

We have explicitly checked that the particle-antiparticle interference term is negligible against the particle-particle interference contribution, $\frac{|\bar n_1(\bold{P},\bold{\Delta P})|^2}{|n_1(\bold{P},\bold{\Delta P})|^2} \lesssim 10^{-2}$. Hence, the last term in the two-particle spectrum $n_2(\bold{P},\bold{\Delta P})$, defined in \Eq{eq:c2}, can be neglected.

In \Fig{fig:n2Conv} we show the two-particle spectrum within our qualitative approximations in out, side, and long direction in the normal phase. This has to be compared to \Fig{fig:n2Moat}, where we show these correlations in the moat regime. In the normal phase, the two-particle spectra decrease monotonically with increasing average pair momentum $|P|$. In the moat regime $n_2$ is a non-monotonic function of $|P|$, with a pronounced peak at nonzero $|P|$ and small relative pair momentum $|\Delta P|$, which continues as a ridge with increasing $|\Delta P|$.
Similar to the single-particle spectrum in \Fig{fig:spec}, the location of this peak is related to the wavenumber of the underlying spatial modulation. The reason is that the average momentum is put on-shell in the particle-particle  interference term \Eq{eq:n1ppapprox}, so the production of particle pairs with relative momenta close to the minimum of the energy $\omega_{\bold{P}_\perp}$, i.e., $|\bold{P}| \approx |\bold{P}_{\rm min}|$, is enhanced. Consequently, also the correlation is enhanced. For the same reason, correlations in the normal phase peak at zero $|\bold{P}|$. The moat regime shifts this peak to nonzero momentum. This is induced by the in-medium modifications of the spectral function of the particles, cf.\ \Eq{eq:s}.

Our result is qualitatively similar to the result in Ref.\ \cite{Pisarski:2021qof}, where such a peak also arises in the two-particle correlation generated through thermodynamic fluctuations. In fact, the underlying reason for the non-monotonic structure of correlations is exactly the same here as in Ref.\ \cite{Pisarski:2021qof}. We also note that the peaks seen in the two-particle correlations are manifestly different from the structures that arise from final-state Coulomb interactions \cite{Sinyukov:1998fc}, since these occur in the relative momentum dependence, while our signal is in the average momentum dependence.

\subsubsection{HBT radii}

To extract the HBT radii, the normalized two-particle correlation $C(\bold{P},\bold{\Delta P})$, defined in \Eq{eq:C}, is considered. At vanishing relative momentum $C(\bold{P},\bold{0}) = 2$. 
This is known as the Goldhaber-Goldhaber-Lee-Pais effect \cite{Goldhaber:1960sf}, and follows trivially from the definition of the interference term, \Eq{eq:n1mixed}. Consequently, the normalized correlations do not show peaks as a function of $|\bold{P}|$ at $|\bold{\Delta P}| = 0$, unlike $n_2$, see \Fig{fig:n2Moat}.
However, ratios of the normalized correlations turn out to be very sensitive to the moat regime. They are shown in \Fig{fig:CratConv} in the normal phase and in \Fig{fig:CratMoat} in the moat regime, and are discussed in \Sec{sec:sum}. 

As as discussed in \Sec{sec:HBTG}, since the emission function connects the spacetime structure of the source to the relative momentum dependence of the correlation, cf.\ \Eq{eq:emf2}, the inverse range of the correlations in $\bold{\Delta P}$ reflects the size of the region of homogeneity in $\bold{x}$. The HBT radii extracted from the range of $C(\bold{P},\bold{\Delta P})$ are interpreted as measuring the spatial extent of the region of homogeneity. We determine the HBT radii through
\begin{align}
R(|\bold{P}|) = \frac{1}{\big|\bold{\Delta P}^*(|\bold{P}|)\big|}\,,
\end{align}
where $\bold{\Delta P}^*$ is the relative momentum where the correlation falls off to half its maximal size, i.e.,
\begin{align}
C\big(\bold{P},\bold{\Delta P}^*\big)=3/2\,,
\end{align}
where we used that $C(\bold{P},\bold{0})=2$.
In \Fig{fig:RHBT} we compare the radii in out, side, and long direction in the moat regime to the normal phase. We see that while the radii in a moat regime are clearly different from the ones in the normal phase, the details depend strongly on the direction of the relative momentum. 
 $R_{\rm out}$ is a bit increased in the moat regime, but most strikingly shows a pronounced dip around $|\bold{P}| \approx |\bold{P}_{\rm min}| = 100$\,MeV. A similar dipping behavior is observed in $R_{\rm side}$, but it is far less pronounced. In this case, the radius beyond the shallow dip is similar to the one in the normal phase. Instead of a dip, $R_{\rm long}$ shows a peak in this region. Thus, the long correlation falls of more rapidly in a moat regime if the average momenta of the particles are close to the minimal one. In summary, also HBT radii are sensitive to the moat regime.

While HBT radii can always be computed in principle, whether they have the desired physical meaning in the present case is questionable. First, we have shown in \Sec{sec:HBTG} that particle-particle interference cannot in general be described by an emission function that allows for an interpretation as phase-space distribution of particles. Second, our correlation functions show a rather strong dependence on the average momentum. Hence, the smoothness approximation which underlies the physical interpretation of HBT radii \cite{Pratt:1997pw} might not be valid here. 

\section{Conclusion}\label{sec:conc}

We have shown that HBT correlations are sensitive probes of a moat regime in heavy-ion collisions. To facilitate this, we developed a field-theoretical formalism that relates $n$-particle spectra to $2n$-point functions of quantum fields on curved hypersurfaces of spacetime. This way, in-medium modifications of particles on hypersurfaces of the expanding medium created in a heavy-ion collision can leave signatures in the spectra of these particles. We have shown that in the Gaussian approximation, these signatures arise in two-particle correlations through the spectral functions of the particles in the medium.

Our qualitative study shows that 
correlation functions of two identical bosons in a moat regime show characteristic peaks at nonzero average pair momentum. This is in contrast to a normal phase without spatial modulations, where these correlations always peak at zero average momentum. This effect can be interpreted as a result of enhanced particle production: Since particles have less energy at the bottom of the moat, they are more abundantly produced carrying the momentum that minimizes their energy, thereby directly enhancing also correlations at this momentum. This has already been pointed out in Ref.\ \cite{Pisarski:2021qof}, where thermodynamic correlations have been analyzed, but a general formalism to describe particle spectra was still lacking. 

We have shown that HBT radii, which are extracted from normalized two-particle correlations, are modified in a moat regime as well.
It turned out that ratios of these normalized two-particle correlations are particularly sensitive probes of a moat regime. Due to their distinguished role in interferometry, normalized correlations of identical bosons, such as pions, are extensively studied in heavy-ion collisions. Their ratios have not been studied yet, but are promising observables to detect a moat regime.

We conclude that momentum-dependent particle correlations can be used as experimental probes of novel phases in QCD in heavy-ion collisions. Based on first indications of a moat regime in the QCD phase diagram found in Ref.\ \cite{Fu:2019hdw}, we expect these signatures to appear in lower-energy collisions, which could be within reach, e.g., of the STAR Fixed-Target program or HADES, and future experiments such as CBM at FAIR.

\begin{acknowledgments}
We are grateful to Eduardo Grossi and Mike Lisa for valuable discussions. Upon completion of this work, we learned that Kenji Fukushima and Yoshimasa Hidaka have done an independent study of HBT in a moat regime. This work is supported by the
Deutsche Forschungsgemeinschaft (DFG, German Research Foundation)
through the Collaborative Research Center TransRegio
CRC-TR 211 ``Strong- interaction matter under extreme conditions'' -- project number
315477589 -- TRR 211. The work of D.H.R.\ is supported 
by the State of Hesse within the Research Cluster
ELEMENTS (Project ID 500/10.006).
\end{acknowledgments}

\begin{appendix}

\section{Particle-antiparticle interference}\label{app:pai}

Here we derive the expression for particle-antiparticle interference, which is the third contribution in \Eq{eq:c2}. We have with the relative and average momenta defined in \Eq{eq:momdef}:
\begin{align}
\begin{split}
&\bar n_1(\bold{p}_\perp,\bold{q}_\perp)\\
&\;= - \frac{1}{2}\lim_{x_\parallel\rightarrow y_\parallel}\int\!d\Sigma^\mu\,d\Sigma'^\nu\,e^{-i\bar p \cdot x}e^{-i\bar q \cdot y}\,(\partial_{\mu}^x+i\bar p_\mu)(\partial_{\nu}^y+i\bar q_\nu)\\
&\;\quad\times \langle \phi(x)\phi(y)\rangle\\
&\;=- \frac{1}{2}\lim_{\Delta X_\parallel \rightarrow 0}\int\!d\Sigma_X\,d\Sigma_{\Delta X}\,e^{-2i\overline{P}\cdot  X}e^{-\frac{i}{2}\overline{\Delta P} \cdot \Delta X}\\
&\;\quad \times\bigg(\frac{1}{4}\big(\partial_{\parallel}^X\big)^2-\big(\partial^{\Delta X}_\parallel \big)^2 +i\overline{\Delta P}_\parallel \partial_{\parallel}^X -i \overline{\Delta P}_\parallel +\frac{1}{4}\overline{\Delta P}_\parallel^2 - \overline{P}_\parallel^2  \bigg)\\
&\;\quad\times\bigg\langle \phi\bigg(X+\frac{1}{2}\Delta X\bigg)\phi\bigg(X-\frac{1}{2}\Delta X\bigg)\bigg\rangle\,.
\end{split}
\end{align}
Carrying out the Wigner transformation yields
\begin{align}\label{eq:as}
\begin{split}
&\bar n_1(\bold{p}_\perp,\bold{q}_\perp)\\
&\;= -\frac{1}{2}\int\!d\Sigma_X\, e^{-2i\overline{P}\cdot  X} \int\!\frac{d\Delta P_\parallel}{2\pi}\, \bigg[ \frac{1}{4}\big(\partial_{\parallel}^X\big)^2 +i\overline{P}_\parallel \partial_{\parallel}^X\\
&\;\quad +\frac{1}{4}\big(\Delta P_\parallel + \overline{\Delta P}_\parallel  \big)^2 - \widetilde P_\parallel^2\bigg]\\
&\;\quad\times\bigg[F\bigg(X,\frac{1}{2}\Delta P\bigg) - \frac{1}{2}\rho\bigg( X,\frac{1}{2}\Delta P \bigg)\bigg]\,.
\end{split}
\end{align}
For this contribution the relative momentum is associated with the relative location and hence enters the Wigner-transformed two-point functions.

Assuming local thermodynamic equilibrium and that gradients in the average location are negligible, we find in analogy to \Eq{eq:s}:
\begin{align}\label{eq:barn}
\begin{split}
&\bar n_1(\bold{p}_\perp,\bold{q}_\perp)\\
&\quad= -\frac{1}{8}\int\!d\Sigma_X\, e^{-2i\overline{P} \cdot X} \int\!\frac{d\Delta P_\parallel}{2\pi}\, \Big[\big(\Delta P_\parallel + \overline{\Delta P}_\parallel  \big)^2\\
&\quad\quad- 4 \overline{P}_\parallel^2 \Big]\; f\bigg(X,\frac{1}{2}\Delta P\bigg)\,\rho\bigg( X,\frac{1}{2}\Delta P \bigg)\,.
\end{split}
\end{align}

\end{appendix}
\hfill
\bibliography{moat}

\end{document}